\documentclass[a4paper,10pt]{article}
\pdfoutput=1 

\usepackage{jheppub} 

\usepackage[T1]{fontenc} 

\usepackage[pass]{geometry}
\newlength\DX \DX=3cm
\paperwidth=\dimexpr\paperwidth-\DX\relax
\hoffset=\dimexpr\hoffset-.5\DX\relax
\newlength\DY \DY=3cm
\paperheight=\dimexpr\paperheight-\DY\relax
\voffset=\dimexpr\voffset-.1\DY-.5\footskip\relax

\usepackage[utf8]{inputenc}
\usepackage[english]{babel}
\usepackage{hyperref}
\usepackage{amsmath}
\usepackage{amssymb}
\usepackage{amsfonts}
\usepackage{amsthm,bm}
\usepackage{physics}
\usepackage{aas_macros}
\usepackage{cleveref}
\usepackage{graphicx}
\usepackage{caption}
\usepackage{subcaption}
\usepackage{leftindex}
\usepackage{multirow}
\usepackage{soul}
\usepackage[normalem]{ulem}

\newcommand{\deriv}[2]{\dfrac{\dd #1}{\dd #2}}

\newcommand{\OO}{\mathcal{O}}

\newcommand{\sch}{Schwarzschild }

\title{Amplitudes and Polarizations of Quadratic Quasi-Normal Modes for a Schwarzschild Black Hole}
\author{Bruno Bucciotti,}
\author{Leonardo Juliano,}
\author{Adrien Kuntz,}
\author{Enrico Trincherini}

\affiliation{Scuola Normale Superiore, Piazza dei Cavalieri 7, 56126, Pisa, Italy and INFN Sezione di Pisa, Largo Pontecorvo 3, 56127 Pisa, Italy}

\affiliation{SISSA, Via Bonomea 265, 34136 Trieste, Italy and INFN Sezione di Trieste and IFPU - Institute for Fundamental Physics of the Universe, Via Beirut 2, 34014 Trieste, Italy}

\emailAdd{bruno.bucciotti@sns.it}
\emailAdd{leonardo.juliano@sns.it}
\emailAdd{adrien.kuntz@sissa.it}
\emailAdd{enrico.trincherini@sns.it}

\abstract{
General Relativity predicts the existence of quadratic quasi-normal modes at second order in perturbation theory. Building on our recent work,
we compute the amplitudes and polarizations of these modes for non-rotating black holes, showing that they are completely determined by the amplitudes and polarizations of linear modes. We obtain the ratio of quadratic to linear amplitudes, which still depends on the initial conditions of the merger through the polarization of linear modes.  However, we demonstrate that this dependence is captured by four fundamental numbers, independent of initial conditions, representing four different combinations of linear modes parities. Additionally, we prove two selection rules regarding the vanishing of classes of quadratic modes. 
Our results are available online as a package which provides the ratio of amplitudes across a broad spectrum of angular momenta.
}

\begin{document}

\maketitle

\section{Introduction}

According to General Relativity (GR), when two black holes merge, the resulting remnant is initially highly deformed but quickly settles into a stationary Kerr black hole. The gravitational wave signal emitted during this last “ringdown” phase is effectively described by black hole perturbation theory~\cite{Regge:1957td,Zerilli:1970aa,1973ApJ...185..635T}.  On the other hand, analytic perturbation theory also plays a key role in the initial phase of the merger, the inspiral phase~\cite{Blanchet:2013haa}. While, for the inspiral, the theory has been pushed to high perturbative orders, the standard paradigm for modeling the ringdown only relies on linear theory.
At linear order, gravitational waves during the ringdown are effectively described by a combination of exponentially damped sinusoids known as quasinormal modes (QNMs)~\cite{Berti:2009kk}.  They are labeled by two angular harmonic numbers $(\ell, m)$, an overtone number $n$ and characterized by a discrete set of complex frequencies $\omega_{\ell mn}$. The real part of $\omega_{\ell mn}$ specifies the QNM oscillation frequency, while the imaginary part determines the decay timescale. In GR the entire linear QNM spectrum is predicted in terms of the mass and angular momentum of the black hole, although the computation of the frequencies cannot be done analytically (see for example \cite{Berti:2009kk,Konoplya:2011qq} for a review on the techniques developed in the literature). The amplitudes of each linear mode at the beginning of the ringdown, on the other hand, are initial conditions that must be fitted using data from merger events or Numerical Relativity (NR) simulations~\cite{Ghosh:2021mrv,LIGOScientific:2016lio,LIGOScientific:2020tif,Buonanno:2006ui,Berti:2007fi,Berti:2007zu,Baibhav:2017jhs,Giesler:2019uxc,Cheung:2023vki}.

Enhanced detector sensitivity over the coming years \cite{Barausse:2020rsu,reitze2019cosmic,punturo:hal-00629986} will enable the capture of more detailed waveforms. These advancements will allow for a deeper analysis of the gravitational waves emitted during the ringdown phase, potentially uncovering second-order perturbation effects in high signal-to-noise ratio events~\cite{Yi:2024elj}. Previous studies \cite{Ma_2022,London_2014,Mitman:2022qdl, Cheung:2022rbm,Cheung:2023vki,Khera:2023oyf,Zhu:2024rej,Redondo-Yuste:2023seq} have already demonstrated that such second-order effects can be identified in certain simulations of binary black hole mergers.   
\medskip

Linear order QNMs obey a homogeneous linear wave equation: a second order differential operator should annihilate the linear metric perturbation. Linearity and homogeneity are the reasons why the QNM amplitudes are initial conditions and they cannot be predicted without modelling the merger. On the contrary, the main output of the analysis is the spectrum of frequencies of the linear QNMs. The frequencies are complex because of the boundary conditions which are outgoing at infinity and infalling at the BH horizon. 

We will now explain second-order perturbation theory, which has been developed for both Schwarzschild and Kerr black holes~\cite{Gleiser:1995gx,Nicasio_1999,Gleiser_2000,Brizuela:2006ne,Brizuela:2007zza,Brizuela:2009qd,Nakano:2007cj,Ioka:2007ak,Ma:2024qcv,Spiers:2023mor,Wardell:2021fyy,Bourg:2024jme,BenAchour:2024skv}, by comparing and contrasting it with the linear theory. The second-order metric perturbations solve a wave equation similar to the first order one: the same second order differential operator now acts on the quadratic perturbations, but this time the equation is not homogeneous. Inhomogeneity comes from a source term, quadratic in the linear perturbations.
For these reasons, the resulting quadratic quasi-normal modes (QQNMs) differ from the linear ones in three aspects, which will be reviewed more fully in the rest of the paper.
\smallskip

First, from the linearity of the differential operator, we can split the quadratic source term into a sum over pairs of linear QNMs. Each pair of linear QNM will couple and give rise to a different QQNM, and the actual quadratic perturbation will be the sum of all the QQNMs thus obtained. The physical interpretation of this fact is that perturbation theory allows us to progressively account for the non-linearities of General Relativity at the perturbative order we consider, and the first interaction is the three graviton vertex.

\begin{figure}
    \centering
    \includegraphics[width=0.8\linewidth]{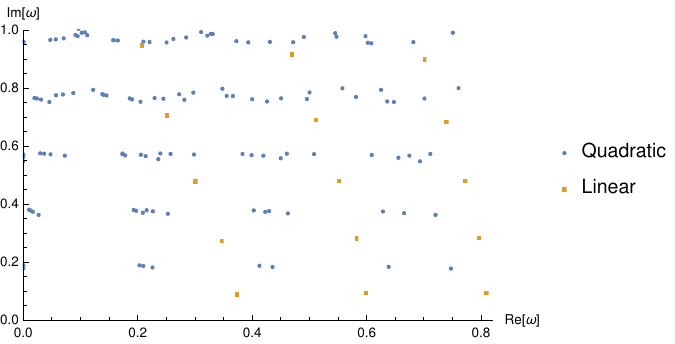}
    \caption{Real and Imaginary part of the frequency of the linear and quadratic QNMs, in units where the BH mass is $M=1$. They represent the QNM oscillation frequency and decay rate respectively. To avoid clutter, we only display quadratic frequencies that can be obtained combining linear modes with $2\le\ell\le4$ and $0\le n\le 4$. Each mode has a mirror mode with negative real part and identical imaginary part.}
    \label{fig:linear_quadratic_frequencies}
\end{figure}

The second point is that, contrary to the linear case, the frequency spectrum at quadratic order is completely fixed by symmetry. Anticipating the discussion in~\cref{sec:linearVSQuad}, if we consider a source term made of two linear QNMs with frequencies $\omega_1$ and $\omega_2$, the frequency of the corresponding QQNM will be either $\omega_1+\omega_2$ or $\omega_1 - \omega_2^*$. Analogously, there is a selection rule on the possible values of the resulting angular momentum.
Since the linear spectrum already contains numerous close-by frequencies, the number of allowed combinations at the quadratic level is huge, as shown in Figure~\ref{fig:linear_quadratic_frequencies}. This figure also highlights that QQNM decay rates can be slower than linear ones (for slightly larger overtone numbers). Still, the total contribution of QQNM to the signal depends on both their decay time and amplitude: if the latter is non-negligible (as for overtones where the amplitude can even be large~\cite{Giesler:2019uxc}), it will be important to include them in future ringdown models.  Therefore, determining the QQNMs amplitudes becomes critical. 
However, the multiplicity of QQNM makes it a challenging task to determine the amplitudes of both linear and quadratic modes by fitting a gravitational wave signal. 

This prompts the discussion of the third and main difference between linear and quadratic QNMs. Because of the source term, QQNMs solve a non-homogeneous equation. By additionally supplying the appropriate QNM boundary conditions, we will show that the amplitude of every quadratic perturbation is, in principle, completely determined in terms of the two linear amplitudes that produce it.
A simple scaling argument shows that the quadratic amplitudes are proportional to the product of the two linear amplitudes sourcing them, thus the most relevant quantity that one can compute is their ratio. Computing these ratio of amplitudes for a non-rotating black hole in GR will be the central goal of this paper.

Because all quadratic amplitudes are fixed, including quadratic modes can improve the fit of the ringdown signal without introducing any new free parameters. This property is highly desirable given the recent debates on the possibility that QNM models can overfit data~\cite{Carullo:2023gtf,Isi:2023nif,Cotesta:2022pci,Baibhav:2023clw}.
\medskip

Nevertheless, computing the amplitude ratios presents several technical challenges due to the redundancy of the metric tensor in encoding the two physical degrees of freedom. The process involves fixing gauge redundancy, selecting master scalar variables to capture the graviton d.o.f., solving  the equations of motion and finally reconstructing the full metric to obtain the gravitational wave amplitude at infinity. On top of this, the second-order master scalars can diverge at infinity~\cite{Gleiser:1995gx,Nicasio_1999,Gleiser_2000,Brizuela:2006ne,Brizuela:2007zza,Brizuela:2009qd,Nakano:2007cj,Ioka:2007ak,Ma:2024qcv}, complicating the extraction of physical effects. 
This is why, so far, only a limited number of second-order amplitudes have been obtained by fitting NR simulations~\cite{Ma_2022,London_2014,Mitman:2022qdl, Cheung:2022rbm,Cheung:2023vki,Khera:2023oyf,Zhu:2024rej} or by integrating BH perturbation theory equations~\cite{Nakano:2007cj,Ioka:2007ak,Bucciotti:2023ets,Redondo-Yuste:2023seq,Ma:2024qcv,Perrone:2023jzq,Bourg:2024jme,BenAchour:2024skv}.  Building on our recent work \cite{Bucciotti:2024zyp}, in this paper we employ two new master scalars that remain regular everywhere, allowing us to derive all the low multipole QQNM amplitudes on a Schwarzschild background after integrating the resulting equations using the Leaver algorithm. Our results are the quadratic counterpart to the well-known tables of linear QNM frequencies that one can find in the literature~\cite{Berti:2009kk}.
Because QQNM seem to only depend weakly on spin~\cite{Cheung:2023vki,Redondo-Yuste:2023seq,Ma:2024qcv,Zhu:2024rej}, we expect our calculation to also be relevant at least for slowly rotating BHs, on top of being expressed in a simpler form than for the Kerr case.
These results are important for two main reasons. Firstly, they can be used to improve the numerical fit of GR to observational data in merger events. Additionally, our computations offer predictions for vacuum non-linear GR effects that can be empirically tested, providing a crucial benchmark against which modified gravity theories or environmental effects can be evaluated.

The present article elaborates on the calculations presented in the companion Letter~\cite{Bucciotti:2024zyp}, while also improving them in two aspects. First, we provide the amplitudes of QQNM for arbitrary waveform polarizations, whereas our previous work focussed on the (physically relevant) case where the waveform is reflection symmetric, implying a circular polarization pattern up to angular factors~\cite{Isi:2021iql}. In particular, we show that the ratio of QQNM to linear QNM amplitudes, previously accepted as a universal quantity independent of the initial conditions of the merger, actually depends on the polarization content of the linear modes, corroborating the recent results in~\cite{Bourg:2024jme}. However, we demonstrate that this dependence is captured by four fundamental numbers, independent of initial conditions, representing four different combinations of parities of linear modes. Thus, QQNM amplitudes still remain completely determined once the amplitudes of the linear modes themselves are known. Second, we prove two selection rules on the vanishing of classes of quadratic modes. All our results are available online~\cite{csvQuadratic}: in particular we provide the complete expression of our quadratic source term, and we provide a convenient Python function that gives the ratio of QQNM to linear QNM amplitudes.
\medskip

The paper is divided into the following sections: \cref{sec:metric_reconstruction_master_scalars} begins with a review of how to extract the two physical degrees of freedom from the metric tensor, both at linear and quadratic orders. We then write down the differential equation governing the evolution of the these two degrees of freedom. In \cref{sec:solving_differential_equation} we get to solving the differential equation, noting that the variables chosen to codify these degrees of freedom, usually called master scalars, are not well suited for our problem. After clarifying the issue, we then perform a master scalar redefinition to resolve it. Our solution is very similar to the one adopted by Nakano and Ioka~\cite{Nakano:2007cj} for the specific case of the dominant QQNM, while it differs from the one chosen by~\cite{Gleiser:1995gx,Nicasio_1999,Gleiser_2000,Brizuela:2006ne,Brizuela:2007zza,Brizuela:2009qd,Ma:2024qcv}. In \cref{sec:physical_observables}, we extract the physical waveform out of our variables, paying a particular attention to the polarization of the QNMs. Finally, section~\ref{sec:final_ratio} contains all our main results. We draw our conclusions in \cref{sec:conclusions}.

\paragraph{Notation} We set $G=c=1$. Our convention for the Fourier transform is
\begin{equation} \label{eq:convFourier}
    \Psi(t) = \int \frac{\mathrm{d} \omega}{2 \pi} e^{-i \omega t} \tilde \Psi(\omega) \; .
\end{equation}
Due to the proliferation of multiple indices in many formula, we will often shorten the notation and write only those indices that cannot be immediately deduced from the context.

\section{Metric reconstruction from Master Scalars}
\label{sec:metric_reconstruction_master_scalars}
In this section we will first decompose the metric tensor fully exploiting the symmetries of the background.
While the metric contains 10 variables, the physical degrees of freedom of a massless spin-two field must be two. We call the variables encoding these degrees of freedom \emph{master scalars}. This section will then review their definition, the differential equation they obey, and how to express the metric in terms of them, following the conventions of Spiers, Pound and Wardell~\cite{Spiers:2023mor}. To simplify the presentation, after an initial discussion that sets up the problem, we will first briefly review linear perturbations, and then extend the discussion to the quadratic ones. Formulas are then presented in a notation that applies to both linear and quadratic modes.
\medskip

We are interested in the small perturbations on top of the Schwarzschild background metric, up to quadratic order. The natural ansatz is
\begin{equation}
    g_{\mu\nu} = \bar g_{\mu\nu} + \varepsilon h^{(1)}_{\mu \nu} + \varepsilon^2 h^{(2)}_{\mu \nu} + \OO(\varepsilon^3) \; ,
\end{equation}
where $\varepsilon$ is the expansion parameter controlling the amplitude of the perturbations and  $\bar g_{\mu\nu}$ is the Schwarzschild metric,
\begin{equation}
    \bar g_{\mu \nu} \mathrm{d}x^\mu \mathrm{d}x^\nu = - f(r) \mathrm{d}t^2 + f^{-1}(r) \mathrm{d}r^2 + r^2 \big( \mathrm{d}\theta^2 + \sin^2(\theta) \mathrm{d}\phi^2 \big) \; ,
\end{equation}
where $f(r) = 1-2M/r$.
Slightly abusing terminology, we will ofter refer to $h_{\mu\nu}$ as \textquotedblleft metric\textquotedblright{} instead of metric perturbation. Indices are raised and lowered using $\bar g_{\mu\nu}$.

We can impose the Einstein field equations in vacuum for $g_{\mu\nu}$: 
\begin{equation} \label{eq:EOMexpanded}
    G_{\mu \nu}[\mathbf{g}] = G_{\mu \nu}^{(0)} + \varepsilon G^{(1)}_{\mu \nu}[\mathbf{h}^{(1)}] + \varepsilon^2 G^{(1)}_{\mu \nu} [\mathbf{h}^{(2)}] + \varepsilon^2 G^{(2)}_{\mu \nu}[\mathbf{h}^{(1)}, \mathbf{h}^{(1)}] + \mathcal{O}(\varepsilon^3) = 0 \; .
\end{equation}
where $G_{\mu \nu}^{(0)}$ is trivially zero because the Schwarzschild metric solves the vacuum field equations, while $G^{(1)}_{\mu \nu}$ and $G^{(2)}_{\mu \nu}$ are respectively linear and bilinear differential operators. 
At order $\varepsilon$ and $\varepsilon^2$ \cref{eq:EOMexpanded} gives
\begin{align}
    G^{(1)}_{\mu \nu} \big[ \mathbf{h}^{(1)} \big] &= 0 \; ,\label{eq:EinstEqFirstOrder} \\
    G^{(1)}_{\mu \nu} \big[ \mathbf{h}^{(2)} \big] &= -  G^{(2)}_{\mu \nu} \big[ \mathbf{h}^{(1)}, \mathbf{h}^{(1)} \big] \equiv S_{\mu\nu}\big[ \mathbf{h}^{(1)}, \mathbf{h}^{(1)} \big]\label{eq:EinstEqSecondOrder} \; .
\end{align}
Since the right hand side of \cref{eq:EinstEqSecondOrder} will \emph{source} the quadratic perturbations, we also call it $S_{\mu\nu}$.

It is convenient to parametrize the 10 components of the metric and the source keeping rotational symmetry in mind.
We will decompose them into ten spherical components: 
\begin{align}
    h_{tt}, \quad h_{tr}, \quad h_{rr}, \quad h_{t+}, \quad h_{r+}, \quad h_{t-}, \quad h_{r-}, \quad h_+, \quad h_{-} \quad \text{and} \quad h_{\circ}\; .\\
    S_{tt}, \quad S_{tr}, \quad S_{rr}, \quad S_{t+}, \quad S_{r+}, \quad S_{t-}, \quad S_{r-}, \quad S_+, \quad S_{-} \quad \text{and} \quad S_{\circ}\; .
\end{align}
Using the tensor spherical harmonics reviewed in \cref{app:tensor_harmonics} we have
\begin{align}
    \label{eq:spherical_decomposition1}
    h_{ab} =& \sum_{\ell,m, \omega} e^{-i \omega t} h_{ab}^{\ell m}Y^{\ell m} \; ,\\
    \label{eq:spherical_decomposition2}
    h_{aB} =& \sum_{\ell,m, \omega} e^{-i \omega t} \big[ h_{a+}^{\ell m} Y_B^{\ell m} + h_{a-}^{\ell m} X_B^{\ell m} \big] \; ,\\
    \label{eq:spherical_decomposition3}
    h_{AB} =& \sum_{\ell,m, \omega} e^{-i \omega t} \big[ h_\circ^{\ell m} \Omega_{AB} Y^{\ell m} + h_+^{\ell m} Y_{AB}^{\ell m} + h_-^{\ell m} X_{AB}^{\ell m} \big] \; ,
\end{align}
and similarly for $S_{\mu \nu}$.
In the previous equation $a,b,\dots$ are indices representing $t,r$, while $A,B,\dots$ represent $\theta,\phi$.
From now on, we will not write ${}^{\ell m}$ above $h$ and $S$ components, leaving it implicit. Due to properties of tensor spherical harmonics, components of the metric with a $+$ subscript are \textit{parity-even} (they do not change under parity), while the $-$ components are \textit{parity-odd} (they take a minus sign under parity). Furthermore, since the background metric $\bar g_{\mu\nu}$ is symmetric under time translations, we focus on perturbations oscillating at a single frequency $e^{-i \omega t}$. We take advantage of the fact that the spectrum is discrete to sum over all QNM frequencies $\omega$. In our conventions, $\omega = \omega^r+i\omega^i$ and $\omega^i<0$. Following~\cite{Dhani:2020nik}, modes with $\omega^r>0$ are called \textquoteleft regular modes\textquoteright{}, while those with $\omega^r<0$ are called \textquoteleft mirror modes\textquoteright. Notice that this definition of mirror modes is different from the one used in~\cite{Bourg:2024jme}, where mirror modes ares defined as modes with $m<0$. However, since for a \sch BH the QNM frequencies do not depend on $m$, we find the former definition more appropriate to our setup. Moreover, mirror modes are also different from retrograde (or counterotating) modes~\cite{Berti:2009kk,Cheung:2023vki}, which are defined as modes with $\text{sign}(\omega^r m) < 0$. 
Finally, we focus on perturbations with $\ell \geq 2$, as according to the peeling theorem~\cite{1961RSPSA.264..309S,1962RSPSA.270..103S,1962JMP.....3..566N} these are the only ones carrying radiation to infinity.

\subsection{Gauge choice}
We will now fix the Regge-Wheeler (RW) gauge, which is the most common choice when performing black hole perturbation theory (see for example~\cite{Spiers:2023mor} for a detailed discussion). 
Metric perturbations transform under diffeomorphisms as
\begin{align}
    h_{\mu \nu}^{(1)} &\rightarrow h_{\mu \nu}^{(1)} + \mathcal{L}_{\xi^{(1)}} \bar g_{\mu \nu} \nonumber \; , \\
    h_{\mu \nu}^{(2)} &\rightarrow  h_{\mu \nu}^{(2)} + \mathcal{L}_{\xi^{(2)}} \bar g_{\mu \nu} + \frac{1}{2} \mathcal{L}^2_{\xi^{(1)}} \bar g_{\mu \nu} + \mathcal{L}_{\xi^{(1)}} h^{(1)}_{\mu \nu} \; ,
    \label{eq:diffeo_on_h}
\end{align}
where $\mathcal{L}_{\xi}$ is the Lie derivative with respect to a vector field $\xi_\mu$.
While the full Einstein tensor is covariant under diffeomorphisms, the operators that appear from its perturbative expansion \cref{eq:EinstEqFirstOrder,eq:EinstEqSecondOrder} have different transformation laws, as emphasised in~\cite{Spiers:2023mor}.
We will denote by $\tilde h^{(i)}$ the metric perturbations taken in Regge-Wheeler gauge, where one imposes
\begin{equation}
    \tilde h_{t+} = \tilde h_{r+} = \tilde h_+ = \tilde h_- = 0 \; .
\end{equation}
Our source term $S_{\mu\nu}$ in \cref{eq:EinstEqSecondOrder} will always be computed using the linearized perturbations in RW gauge, without adding the tilde in order to improve readability.

By applying a generic infinitesimal diffeomorphism as in \cref{eq:diffeo_on_h}, we can find the relation between the metric in Regge-Wheeler gauge and in a generic gauge. In \cref{sec:physical_observables} we will do this to compute the metric in asymptotically flat gauge, where the physical waveform is most easily extracted.

\subsection{Main formulas} \label{sec:mainFormulas}
We now present the formulas in a language which unifies the discussion of linear and quadratic modes. For each parity sector all the $\tilde h^{(1)}$ variables can be expressed in terms of a single master scalar, which we take to be the well known Cunningham-Price-Moncrief $\psi_{-}^{(1)}$ for the odd sector, and the Zerilli-Moncrief $\psi_{+}^{(1)}$ for the even sector (see \cite{Hui:2020xxx} for an interesting strategy to define them constructively). The master scalars are defined as
\begin{align}
    \label{eq:master_scalar_even}
    \psi_{+}^{(i)} = \frac{2r}{\lambda_{1}^2}\left[ r^{-2}\tilde{h}_\circ^{(i)} +\frac{2}{\Lambda(r)} \left( f^2\tilde{h}_{rr}^{(i)} - rf(r^{-2}\tilde{h}_\circ^{(i)})' \right) \right] \; ,\\
    \label{eq:master_scalar_odd}
    \psi_{-}^{(i)} = \frac{2r}{\mu^2}\left[ (\tilde{h}_{t-}^{(i)})' + \frac{M}{r^2 f(r)} (\tilde{h}_{t -}^{(i)}-\tilde{h}_{r -}^{(i)}) +i \omega \tilde{h}_{r-}^{(i)} - \frac{2}{r}\tilde{h}_{t-}^{(i)} \right] \; ,
\end{align}
and they obey the Zerilli and Regge-Wheeler equations, obtained by imposing the equations of motion~\eqref{eq:EinstEqFirstOrder}-\eqref{eq:EinstEqSecondOrder}
\begin{align}
    \label{eq:RWZ_equation}
    \deriv{\psi_{\pm}^{(i)}}{r_*^2}+\omega^2 \psi_{\pm}^{(i)}-V_{\pm}(r) \psi_{\pm}^{(i)} = S^{(i)}_\pm \; ,
\end{align}
where $'=\mathrm{d}/\mathrm{d}r$, $r_* = r+2M \ln\left(\frac{r}{2M}-1\right)$ is the tortoise coordinate and $V_{+}(r)$, $V_{-}(r)$ are the Zerilli and Regge-Wheeler potentials
\begin{equation}
    \begin{split}
        V_+(r) &:= \frac{f(r)}{\Lambda(r)^2}\left[ \frac{\mu^4}{r^2} \left( \lambda_{1}^2 + \frac{6M}{r}\right) + \frac{36M^2}{r^4} \left( \mu^2 + \frac{2M}{r} \right) \right] \; , \\
        V_-(r) &:= f(r)\left(\frac{\ell(\ell+1)}{r^2} - \frac{6M}{r^3}\right) \label{eq:RWpot} \; .
    \end{split}
\end{equation}
Since there is no source at linear order $S^{(1)}=0$, we will simplify notation and rename $S^{(2)}\equiv S$. We will explain in more details in the next section how one can obtain the source term in eq.~\eqref{eq:RWZ_equation} from Einstein's equations at second order~\eqref{eq:EinstEqSecondOrder}, and also highlight some properties of the source term. However, since its explicit expression is quite involved, we give it only in the companion \textsc{Mathematica} files~\cite{csvQuadratic}.
Following \cite{Spiers:2023mor} we also defined
\begin{equation}
\label{eq:def:lambda_mu_Lambda}
    \lambda_s = \sqrt{\frac{(\ell +|s|)!}{(\ell -|s|)!}},\quad \mu^2 = \lambda_{1}^2-2 = (\ell+2)(\ell-1),\quad \Lambda(r) = \mu^2+\frac{6M}{r} \; .
\end{equation}
There is a one-to-one relation between master scalars and the metric perturbations they describe: indeed
the latter can be reconstructed from the former as
\begin{equation} \label{eq:metric_reconstruction}
    \begin{split}  
    \tilde h_{tt}^{(i)} &= f(r)^2 \tilde h_{rr}^{(i)} + 2 f(r) S_+^{(i)}  \; , \\
    \tilde h_{tr}^{(i)} &= -i \omega r \psi_+^{(i)} {}' + \frac{-i \omega}{f(r) \Lambda(r)} \left [ \mu^2 \left(1-\frac{3M}{r} \right) - \frac{6M^2}{r^2}\right] \psi_+^{(i)} + \frac{2r^2}{\lambda_1^2} \left(S_{tr}^{(i)} +i \omega \frac{2r}{\Lambda(r)f(r)} S_{tt}^{(i)} \right) \; ,  \\
    \tilde h_{rr}^{(i)} &= \frac{1}{4r^2 f(r)^2} \left[\Lambda(r) \left( \lambda_1^2 r \psi_+^{(i)} -2\tilde h_{\circ}^{(i)} \right) + 4r^3 f(r) \left(r^{-2} \tilde h_{\circ}^{(i)} \right)'   \right] \; , \\
    \tilde h_{t-}^{(i)} &= \frac{1}{2}f(r)\left(r \psi_-^{(i)} \right)' + \frac{2r^2}{\mu^2}S_{t-}^{(i)} \; , \\ 
    \tilde h_{r-}^{(i)} &= \frac{i \omega}{2f(r)} r \psi_-^{(i)} + \frac{2r^2}{\mu^2} S_{r-}^{(i)} \; , \\ 
    \tilde h_{\circ}^{(i)} &= r^2 f(r) \psi_+^{(i)} {}' + \frac{r}{2\Lambda(r)}\left[\lambda_2^2 + \frac{6M}{r}\left( \mu^2 + \frac{2M}{r}\right) \right]\psi_+^{(i)} - \frac{4r^4}{\lambda_1^2 \Lambda(r)}S_{tt}^{(i)} \; ,
    \end{split}
\end{equation}
where $S_{\mu\nu}=0$ at linear order, and the frequency $\omega$ is the one of the mode under consideration.

\subsection{Linear versus quadratic perturbations}\label{sec:linearVSQuad}
The goal of this subsection is to highlight the similarities and differences between linear and quadratic perturbations.
At first order,
the Regge-Wheeler and Zerilli equations are linear homogeneous differential equations, so the amplitude of the master scalar can be rescaled at will and indeed in our setup the linear waveform amplitude is determined by fitting to data (either interferometers signals or Numerical Relativity simulations). The main prediction of the differential equations come when we supplement them with outgoing boundary conditions at infinity and infalling at the black hole horizon, which are the boundary conditions for QNM: the spectrum of allowed frequencies becomes discrete~\cite{1975RSPSA.344..441C}. Modes are thus labeled by an overtone number $n$. Even for a single overtone $n$, there are two solutions to the RW and Zerilli equations~\eqref{eq:RWZ_equation}, one with positive real part of the frequency which we label as $\omega_{\ell n +}$ (called regular mode), and one with negative real part given by $\omega_{\ell n -} = - (\omega_{\ell n +})^*$, where $*$ denotes complex conjugation. Thus, in general, linear modes are indexed by the supplementary mode number $\mathcal{N}=(n, \mathfrak m)$ where $\mathfrak m = \pm$ denotes a regular or mirror mode. 

As we will see, while much of the formulas in section~\ref{sec:mainFormulas} are only slightly modified at second order, the Regge-Wheeler and Zerilli equations now predict the amplitude of the quadratic modes given the amplitude of the linear ones, while the set of allowed frequencies is trivial to obtain.
As one can see from~\cref{eq:EinstEqSecondOrder}, schematically we have on the left-hand side the same differential operator $G^{(1)}$ as at linear order, now acting on the second order perturbation. The main difference is the appearance of a source term, completely determined once the linear perturbations are fixed. We will still demand QNM boundary conditions on the solutions of \cref{eq:EinstEqSecondOrder}.
Given the (bi)linearity properties of $G^{(i)}$, we can look for second-order solutions of \cref{eq:EinstEqSecondOrder} by considering two fixed linear perturbations in the source term, say described by $\psi_1^{(1)}$ and $\psi_2^{(1)}$, and then constructing the full solution using a superposition. While the linear perturbations can be generic at this point, we will take them to be linear QNMs characterized by $(\ell_1,m_1,n_1,\mathfrak m_1,p_1),\;(\ell_2,m_2,n_2,\mathfrak m_2,p_2)$ (with frequencies $\omega_1$ and $\omega_2$) where $p$ is the parity, even for Zerilli and odd for Regge-Wheeler. 
We will just assume here that modes $1$ and $2$ are not the same and add up in the total waveform, while introducing later on a symmetry factor in order to cover the case where  $(\ell_1,m_1,n_1,\mathfrak m_1,p_1) = (\ell_2,m_2,n_2,\mathfrak m_2,p_2)$ (see also the toy model in appendix~\ref{app:toy_model} for a pedagogical introduction to this symmetry factor).

Given the symmetries of the background, the source term of \cref{eq:EinstEqSecondOrder} is very constrained by selection rules. For example, the only possible time dependence of the source is (without any loss of generality, we take $\omega^r_1\ge\omega^r_2>0$ here)
\begin{equation}
    e^{\omega_1^i t}\cos(\omega_1^r t) \times e^{\omega_2^i t} \cos(\omega_2^r t+\varphi_0) \propto e^{(\omega_1^i+\omega_2^i) t}\left[\cos((\omega_1^r-\omega_2^r)t-\varphi_0) + \cos((\omega_1^r+\omega_2^r)t+\varphi_0)\right] \; ,
\end{equation}
so that at quadratic order $\omega = \omega_1+\omega_2$ or $\omega = \omega_1-\omega_2^*$ or their mirror modes at second order $\omega=-\omega_1^*-\omega_2^*$, $\omega=-\omega_1^*+\omega_2$. Notice that, due to the complicated way in which time derivatives appear in the source term, the actual amplitudes of the two cosines in the source are not equal but  generically unrelated.
Similarly, we have selection rules on the angular momentum
\begin{equation}
    \ell = |\ell_1-\ell_2|, \dots, \ell_1+\ell_2,\qquad
    m = m_1+m_2 \; .
\end{equation}
In the actual source term, selection rules are imposed by the fact that the source is proportional to the $3j$ symbol $\begin{pmatrix}
        \ell_1&\ell_2&\ell\\
        m_1&m_2&-m
    \end{pmatrix}$. In fact, the only dependence of the source term on $m$, $m_1$ and $m_2$ is fully contained in the $3j$ symbol because of angular momentum conservation (see later on section~\ref{sec:theorem}).

Finally, parity imposes
\begin{equation}
    \label{eq:selection_rule_parity}
    (-1)^{\ell_1+ p_1} \times (-1)^{\ell_2+ p_2} = (-1)^{\ell+ p}\Rightarrow (-1)^{\ell_1+\ell_2+\ell} = (-1)^{p_1+p_2+p} \; ,
\end{equation}
where the parity of a mode is determined by its intrinsic parity $p$, which is $+1$ for the Zerilli sector and $-1$ for the Regge-Wheeler sector, and by the parity of the spherical harmonics $(-1)^\ell$.

The most general solution of \cref{eq:EinstEqSecondOrder} will be a combination of a particular solution plus the most general solution of the homogeneous equation. However the homogeneous equation is identical to \cref{eq:EinstEqFirstOrder}, so imposing QNM boundary conditions on a homogeneous solution will select the linear QNM frequencies. The linear amplitudes are then artificially shifted by $\OO(\epsilon)$, but there is no physics in this amplitude redefinition in our approach since linear amplitudes just get fitted to data~\footnote{On the other hand, it would be important to take into account this shift if we were trying to relate amplitudes to the initial conditions before merger, as e.g. in~\cite{PhysRevD.34.384}}. On the other hand, new solutions are allowed at frequencies $\omega_1+\omega_2$ and $\omega_1-\omega_2^*$ and their mirrors at second order, because of the non-trivial source term. For these frequencies, we will see that the freedom to add an arbitrary solution of the homogeneous equation will allow us to always find a non-trivial solution with QNM boundary conditions. Thus we recover a well-known result: the frequencies at quadratic order are exactly those appearing in the source term.

Despite these distinctions, the redundancies of $h^{(2)}_{\mu\nu}$ and the consequent reduction to two master scalars containing the graviton polarizations is very similar to the linear order. We specialize $h^{(2)}$ to a given parity, frequency $\omega$ and angular momentum $\ell,m$, compatible with the selection rules described above. Keeping the same definitions for the master scalars (\cref{eq:master_scalar_even,eq:master_scalar_odd}), one discovers that the Regge-Wheeler and Zerilli equations are modified by a source term $S(r)$ as in~\cref{eq:RWZ_equation}. The metric reconstruction also contains the components of $S_{\mu\nu}$ besides the quadratic master scalars (\cref{eq:metric_reconstruction}).

\section{Solving the differential equation}
\label{sec:solving_differential_equation}
In the previous section, we argued that the set of differential equations~(\ref{eq:EinstEqSecondOrder}) can be reduced to a single Regge-Wheeler or Zerilli equation with a known source term \cref{eq:RWZ_equation}, which we now aim to solve. However, we will encounter a technical issue: while $\psi^{(1)}_\pm$ approaches $\text{const.}\times e^{\pm i\omega r_*}$ at both infinity and the horizon, $\psi^{(2)}_\pm$ will a priori be more divergent. This section is dedicated to understanding the problems that arise when this happens and how to avoid them using new master scalars. Although the problem is rather technical in nature, it is possible to understand some of its features through a toy-model, which we present in appendix~\ref{app:toy_model}.
Finally, we will extend the Leaver method to find solutions of the RW/Z equation with a source term and satisfying QNM boundary conditions. In this section we will drop the superscript ${}^{(2)}$ from $\psi^{(2)}$, leaving it implicit.
\medskip

We are interested in solving \cref{eq:RWZ_equation} with a given $S(r)$.
The variation of constants method is a well known way of solving linear non-homogeneous equations, and it leads to $\psi(r) = \psi_p(r) + \psi_h(r)$ where $\psi_p(r)$ is a particular solution of the non-homogeneous equation, while $\psi_h(r)$ is a generic solution of the associated homogeneous equation. While we know from the analysis at linear order that $\psi_h(r)$ behaves like $e^{\pm i \omega r_*}$ as $|r_*|\rightarrow\infty$, no assumption can be made a priori about $\psi_p(r)$.

The reason why a divergent $\psi_p(r)$ would not contradict the regularity expected of the physical waveform is that $S_{\mu\nu}$ comes into play when reconstructing the metric perturbation $\tilde h_{\mu\nu}^{(2)}$, contrary to what happens at linear order where $S_{\mu\nu}^{(1)}\equiv 0$. The divergences of $\psi$ can then cancel against those of $S_{\mu\nu}$. 
While this observation avoids a logical contradiction, the divergent behaviour of $\psi(r)$ at the horizon and at infinity is not free of difficulties.

\subsection{Divergent Master Scalars}
Let us assume that at infinity and at the horizon $\psi(r)\sim e^{\pm i\omega r_*}\times P_\pm(r)$ as $r_*\rightarrow \pm \infty$ respectively, where $P_+(r)$ is a power series in $1/r$ and in $P_-$ is in $(r-2M)$.
Divergences appear whenever the power series $P_\pm(r)$ contain poles. In addition, let $c_1,\,c_2$ be the coefficients of the linear combination defining $\psi_h(r)$.
\medskip

We can point to three main issues that arise when $\psi(r)$ displays a divergent behaviour at infinity or the horizon. While they depend on the precise way in which we are approaching the problem, it is worth noting that similar difficulties arise in alternative approaches too~\cite{Bucciotti:2023ets,Brizuela:2009qd}.
\smallskip

First, ensuring the QNM boundary conditions is trivial for $\tilde h_{\mu\nu}^{(1)}$, because one simply has to select the solution $\psi^{(1)}(r)$ that goes to $e^{\pm i\omega r_*}$ at $r_*\rightarrow\pm\infty$ respectively. For quadratic perturbations, this is far from trivial. The reason is that $\psi_h(r)$ should be chosen so that $\tilde h_{\mu\nu}^{(2)}$ satisfies the QNM boundary conditions, but choosing different $c_1,c_2$ only modifies the non-singular terms of $P_\pm(r)$ (because solutions of the homogeneous equation are regular). The conclusion is that checking that the leading (divergent) behaviour of $\psi(r)$ is outgoing at infinity and infalling at the horizon is not enough: we need control of the subleading terms up to the first regular one.

Similarly, the amplitude of $\tilde h_{\mu\nu}^{(2)}$ cannot be independent of $c_1,c_2$. Thus even our main object of interest, the amplitude, is sensitive to the subleading terms in $\psi(r)$. All the divergent terms in $\psi(r)$ are due to a poor choice of variables, and they all cancel against the divergences of $S_{\mu\nu}$ in \cref{eq:metric_reconstruction}. The amplitude of $\tilde h_{\mu\nu}^{(2)}$ is actually critically determined by the first regular term of $\psi(r)$, which would be very hard to extract if we were to numerically integrate \cref{eq:RWZ_equation}.

Lastly, we could rephrase our whole computation accounting for generic order $\epsilon$ by describing metric perturbations as $g_{\mu\nu}=\bar g_{\mu\nu}+\epsilon \mathbf h_{\mu\nu},\;\mathbf h_{\mu\nu}=h_{\mu\nu}^{(1)}+\epsilon h_{\mu\nu}^{(2)}+\OO(\epsilon^2)$. We would then define the master scalars $\phi_\pm = \psi_\pm^{(1)}+\epsilon \psi_\pm^{(2)}+\OO(\epsilon^2)$ according to \cref{eq:master_scalar_even,eq:master_scalar_odd} at all orders, and they would obey
\begin{equation}
    \deriv{\phi_{\pm}}{r_*^2}+\omega^2 \phi_{\pm}-V_{\pm}(r) \phi_{\pm} = \epsilon\, \mathbf S(r), \quad \mathbf S(r) = S(r) + \OO(\epsilon) \; .
\end{equation}
Our problem would then be reduced to solving for the master scalars perturbatively in $\epsilon$, employing a single $\mathbf h_{\mu\nu}$.
However this way of describing the problem seems to require $|\psi_\pm^{(1)}|\gg \epsilon |\psi_\pm^{(2)}|$ for \emph{all} $r$, which is contradicted at large $r_*$ if we find that $\psi_\pm^{(2)}$ is more divergent that $\psi_\pm^{(1)}$.

Actually this last problem will turn out to be fictitious, because instead of expanding the master scalars in powers of $\epsilon$ we chose to expand the metric perturbations. We will later verify that all the divergences of the master scalars cancel when reconstructing the metric perturbations, thus validating our $\epsilon$ expansion.

\subsection{Bypassing the problem}
As we explained in the previous subsection, we cannot easily work with the master scalars that we defined. This problem has already been noticed in the literature, and was addressed in \cite{Gleiser:1995gx,Nicasio_1999,Gleiser_2000,Nakano:2007cj,Brizuela:2006ne,Brizuela:2007zza,Brizuela:2009qd,Silva:2023cer}. Among these works only~\cite{Nakano:2007cj} was targeted towards finding QNM solutions, although only for the dominant quadratic QNM, while the method employed in other works was not directly applicable to our setting. 
In this subsection we thus adapt their main ideas to our problem.

The crux of the previous discussion is that the divergences of $\psi(r)$ are spurious and should cancel against those of $S_{\mu\nu}$ in \cref{eq:metric_reconstruction}. The idea is then to analyze the (opposite) divergences of $S_{\mu\nu}$ and move them into the master scalar, redefining and regularizing it.
It is not clear from \cref{eq:metric_reconstruction} that this should be doable, because of the intricate way in which $\psi$ and $S_{\mu\nu}$ present themselves. However we can work directly at the level of \cref{eq:RWZ_equation}: we will study the divergences of $S(r)$ and redefine $\psi(r)$ accordingly.

We define the \emph{regular} master scalar
\begin{equation}
    \label{eq:regulated_master_scalar}
    \Psi(r) = \psi(r) + \Delta(r) \psi_1^{(1)}\psi_2^{(1)} \; ,
\end{equation}
where $\Delta(r)$ is a yet-to-be-determined function which can depend on the frequency, the angular momentum and the parity of each of the three modes, which we leave implicit from now on. Substituting \cref{eq:regulated_master_scalar} into \cref{eq:RWZ_equation}, we get that the regulated master scalar satisfies
\begin{equation}
    \label{eq:regulated_RWZ_equation}
    \deriv{\Psi}{r_*^2}+\omega^2 \Psi-V(r) \Psi = \mathfrak S(r),\quad
    \mathfrak S(r) = S(r) + \left[\deriv{ }{r_*^2}+\omega^2-V(r)\right]\left(\Delta(r) \psi_1^{(1)}\psi_2^{(1)}\right) \; ,
\end{equation}
which is a Regge-Wheeler or Zerilli differential equation with a modified source term.
Moreover, we can factor out the linear modes by defining
\begin{equation} \label{eq:sourceSs}
    S(r) = s(r) \psi_1^{(1)}\psi_2^{(1)},\quad
    \mathfrak S(r) = \mathfrak s(r) \psi_1^{(1)}\psi_2^{(1)} \; ,
\end{equation}
so that $s(r),\,\mathfrak s(r)$ are independent of the linear amplitudes. Such a factorization implicitly relies on the fact that the source term is symmetric if we exchange the two perturbations $\psi_1$ and $\psi_2$, which is of course satisfied by the true source term by virtue of the properties of $G^{(2)}$ in eq.~\eqref{eq:EinstEqSecondOrder}.
As discussed in \cite{Nakano:2007cj,Bucciotti:2023ets} the condition for the regularity of $\Psi(r)$ is that $\mathfrak s(r)$ should vanish like $(r-2M)$ at the horizon and as $1/r^2$ at infinity.

We highlight that, due to its definition, $\mathfrak S(r)$ contains first or second order $r$ derivatives of the linear modes. While we can get rid of the second derivatives using the linear Regge-Wheeler and Zerilli equations, the first derivatives cannot be eliminated, hence $s(r)$ and $\mathfrak s(r)$ will contain ratios of the form $\psi_i^{(1)}{}'/\psi_i^{(1)}$. Despite this, our main concern will be to control the asymptotic behaviour of $\mathfrak S(r)$ at the horizon and at infinity, so that these ratios can be computed order by order in $(r-2M)$ and $1/r$ (see~\cref{app:code}). Later, after having defined the regular master scalars, we will have to work with $\mathfrak S(r)$ at finite $r$: at that point we will keep the first derivatives of $\psi^{(1)}_{1,2}$ explicit, writing
\begin{align} \label{eq:Sreg}
    \mathfrak S(r) &= \mathcal{F}_1(r)\psi^{(1)}_1\psi^{(1)}_2+\mathcal{F}_2(r) \big( \psi^{(1)}_1{}'\psi^{(1)}_2 + \psi^{(1)}_1{}\psi^{(1)}_2{}' \big) \nonumber \\
    &+\mathcal{F}_3(r)\big(\psi^{(1)}_1{}'\psi^{(1)}_2-\psi^{(1)}_1\psi^{(1)}_2{}'\big)+\mathcal{F}_4(r)\psi^{(1)}_1{}'\psi^{(1)}_2{}' \; ,\\
    \mathfrak s(r) &= \mathcal{F}_1(r)+\mathcal{F}_2(r)\big(\frac{\psi^{(1)}_1{}'}{\psi^{(1)}_1}+\frac{\psi^{(1)}_2{}'}{\psi^{(1)}_2} \big) +\mathcal{F}_3(r)\big(\frac{\psi^{(1)}_1{}'}{\psi^{(1)}_1}-\frac{\psi^{(1)}_2{}'}{\psi^{(1)}_2} \big)+\mathcal{F}_4(r)\frac{\psi^{(1)}_1{}'}{\psi^{(1)}_1}\frac{\psi^{(1)}_2{}'}{\psi^{(1)}_2} \; .
\end{align}
where we give the explicit expression of the $\mathcal{F}_i$ functions in the companion files~\cite{csvQuadratic}.

We discover by inspection that $s(r)$ is always already regular at the horizon, but not so at infinity. Thus we take a polynomial ansatz for $\Delta(r)$
\begin{equation}
    \label{eq:delta_definition}
    \Delta(r) = a_2 r^2 + a_1 r \; .
\end{equation}
Our code (see~\cref{app:code}) computes the required $\{a_1,a_2\}$ and therefore $\mathfrak s(r)$. 

We draw the reader's attention to the fact that $\Delta(r)$ can contain $1/\omega$ terms. Such terms in Fourier space should be interpreted as time integrals in time domain, meaning that generically there is no master scalar redefinition that regulates both the horizon and infinity while also being local in time. However, there is nothing conceptually difficult about this, and the technical complications of the time integrals are completely absent in Fourier space. The need for such integrals was sidestepped by Ioka and Nakano~\cite{Nakano:2007cj}, who work in time domain, by effectively working with the time derivative of $\psi(r)$ and then performing master scalar redefinitions which are local in time. The (minor) drawback of this approach is that the so-defined master scalar only allows one to reconstruct the time derivatives of the metric perturbations.

Let us conclude this section by commenting on the arbitrariness of the regulation procedure. Indeed, one could have defined other regular master scalars by including a constant term in the regulator~\eqref{eq:delta_definition}, and they would have displayed well-behaved asymptotics as well. However, this is just a matter of definitions, and the asymptotic metric that we will compute in section~\ref{sec:physical_observables} will not depend on this arbitrary choice. This arbitrariness, however, prevents us from making a direct comparison with the work of Nakano and Ioka~\cite{Nakano:2007cj}, which provides the value of the quadratic ratio only for the RW and Zerilli variables and for the dominant QQNM. Our strategy to regulate the master scalars is in some sense \textquotedblleft minimal\textquotedblright{} because we only subtract the divergent terms without affecting the regular ones.

\subsection{Extension of Leaver method} \label{sec:leaver}

The problem of finding the amplitude of quadratic modes has now been reduced to finding a solution to the RW and Zerilli equations at second order with appropriate boundary conditions:
\begin{align} \label{eq:BC2nd}
    \Psi &= M \mathcal{A} e^{i \omega r_*} + \mathcal{O}\bigg( \frac{1}{r} \bigg) \; , \quad \text{for } r \rightarrow \infty \; , \\
    \Psi &= M \mathcal{B} e^{-i \omega r_*} + \mathcal{O}\bigg( 1- \frac{2 M}{r} \bigg) \; , \quad \text{for } r \rightarrow 2M \label{eq:BC2ndH} \; ,
\end{align}
where $\mathcal{A}$ and $\mathcal{B}$ are constants (the $M$ factor is added to have dimensionless amplitudes), and $\omega = \omega_1 + \omega_2$ is the frequency of the nonlinear modes.
Although several analytic or numerical methods can be used both at first and second order to compute the wavefunction from the RW and Zerilli equations~\cite{PhysRevD.35.3621,Iyer:1986vv,1985ApJ...291L..33S,2003PhRvD..68b4018K,Matyjasek_2017,Hatsuda:2023geo,motl2003asymptotic,Ansorg:2016ztf, Ripley:2022ypi,Aminov:2020yma,Bonelli:2021uvf,Bonelli:2022ten,Aminov:2023jve,Nakano:2007cj,Bucciotti:2023ets}, we choose here to use one of the most accurate numerical method: Leaver's algorithm~\cite{leaver}. The application of Leaver method to the second-order modes has been streamlined in~\cite{Nakano:2007cj}, but we will review the procedure here for completeness.

Leaver's method only applies to the RW equation, while we need to solve both RW and Zerilli equations. However, it is easy to transform the Zerilli equation at second order to a RW form by defining the Chandrasekhar variable $\chi^{(2)}$~\cite{1975RSPSA.344..441C}:
\begin{equation} \label{eq:ChandTransf2nd}
    \chi^{(2)} = \frac{1}{\mu^2(\mu^2+2)/6-2iM\omega} \bigg[  \bigg(\frac{1}{6} \mu^2(\mu^2+2) + \frac{12M^2(r-2M)}{r^2(\mu^2 r+6M)} \bigg) \Psi - 2M \frac{\mathrm{d} \Psi}{\mathrm{d}r_*} \bigg] \; ,
\end{equation}
where we recall that $\mu^2 = (\ell+2)(\ell-1)$. 
If $\Psi$ obeys a Zerilli equation with a source term of the form~\eqref{eq:Sreg}, then using the RW and Zerilli equations for both linear and second-order modes it is straightforward to show that $\chi^{(2)}$ obeys a RW equation with another source term of the form
\begin{align} \label{eq:RWeqChandra}
    &\frac{\mathrm{d}^2 \chi^{(2)}}{\mathrm{d}r_*^2} + \big( \omega^2 - V_- \big) \chi^{(2)} = \bar{\mathfrak S} \; , \\
    \bar{\mathfrak S} &= \mathcal{G}_1 \psi_{1}^{(1)} \psi_{2}^{(1)} + \mathcal{G}_2 \big( (\psi_{1}^{(1)})' \psi_{2}^{(1)} + \psi_{1}^{(1)} (\psi_{2}^{(1)})' \big) \nonumber \\
    &+ \mathcal{G}_3 \big( (\psi_{1}^{(1)})' \psi_{2}^{(1)} - \psi_{1}^{(1)} (\psi_{2}^{(1)})' \big) + \mathcal{G}_4 (\psi_{1}^{(1)})' (\psi_{2}^{(1)})' \; ,
\end{align}
where $V_-$ is the RW potential given in eq.~\eqref{eq:RWpot}, and we give the explicit expression of the $\mathcal{G}_i$ functions in the companion \textsc{Mathematica} notebook~\cite{csvQuadratic}. 
Moreover, $\chi^{(2)}$ also satisfies boundary conditions of the kind~\eqref{eq:BC2nd}-\eqref{eq:BC2ndH}, and the transformation~\eqref{eq:ChandTransf2nd} is normalized in such a way that the asymptotic amplitudes of $\chi^{(2)}$ and $\Psi$ coincide for $r \rightarrow \infty$. Similarly, the Chandrasekhar transformation can be used to obtain the Zerilli solution at linear order when we will need it, using the equation
\begin{equation}
    \chi^{(1)}_{i} = \frac{1}{\mu^2_i(\mu^2_i+2)/6+2iM\omega_i} \bigg[  \bigg(\frac{1}{6} \mu^2_i(\mu^2_i+2) + \frac{12M^2(r-2M)}{r^2(\mu^2_i r+6M)} \bigg) \psi^{(1)}_{i} + 2M \frac{\mathrm{d} \psi^{(1)}_{i}}{\mathrm{d}r_*} \bigg] \; ,
\end{equation}
where $i=1,2$. Once again, if $\psi_i^{(1)}$ solves the homogeneous RW equation at first order, then $\chi_i^{(1)}$ solves the homogeneous Zerilli equation and has the same asymptotic amplitude than $\psi_i^{(1)}$ for $r \rightarrow \infty$. 

Let us now describe the Leaver algorithm for solving RW equation at first and second order. At first order, we make an ansatz for the RW variable of the form ($i=1,2$):
\begin{equation}
    \psi_{i}^{(1)} = M A_i(r) \sum_{n=0}^{\infty} a_{n}^{(1)} \bigg( 1 - \frac{2M}{r} \bigg)^n \; ,
\end{equation}
where 
\begin{align}
    A_i(r) &= \bigg( \frac{r}{2M}-1 \bigg)^{\rho_i} \bigg( \frac{r}{2M} \bigg)^{-2 \rho_i} \exp{ \bigg[-\rho_i \bigg( \frac{r}{2M} - 1\bigg) \bigg]} \; , \\
    \rho_i &= -2i M \omega_i \; .
\end{align}
By factoring out the function $A_i(r)$ in the ansatz, one can obtain a wavefunction which satisfies the boundary conditions~\eqref{eq:BC2nd} if the series $\sum_n a_{n}^{(1)}$ converges. The RW equation is then obeyed provided the coefficients $a_{n}^{(1)}$ satisfy the recursion relation:
\begin{align}
    \alpha^{(1)}_0 a_{1}^{(1)} + \beta^{(1)}_0 a_0^{(1)} &= 0 \; , \\
    \alpha^{(1)}_n a_{n+1}^{(1)} + \beta^{(1)}_n a_n^{(1)} + \gamma^{(1)}_n a_{n-1}^{(1)} &= 0 \; ,
\end{align}
where $a_0$ is arbitrary and represents the freedom in the choice of normalization of the linear RW variable, and the coefficients $\alpha_n^{(1)}$, $\beta_n^{(1)}$ and $\gamma_n^{(1)}$ are given by
\begin{align}
    \alpha_n^{(1)} &= n^2 +(2 \rho_i+2) n + 2 \rho_i+1 \; , \\
    \beta_n^{(1)} &= - \big( 2n^2 + (8 \rho_i+2) n + 8 \rho_i^2 + 4\rho_i + \mu^2 - 1\big) \; , \\
    \gamma_n^{(1)} &= n^2 + 4 \rho_i n + 4 \rho_i^2 - 4 \; .
\end{align}
The requirement of convergence of $\sum_n a_{n}^{(1)}$ provides the value of $\omega$ for linear modes. We numerically obtain it using a simple minimization algorithm. 

At second order, the ansatz for the RW variable is changed to
\begin{equation}
    \Psi = M A_1(r) A_2(r) \sum_{n=0}^{\infty} a_{n}^{(2)} \bigg( 1 - \frac{2M}{r} \bigg)^n \; ,
\end{equation}
and similarly for the Chandrasekhar transformation of the Zerilli variable $\chi^{(2)}$. Plugging the ansatz in the equations of motion (either~\eqref{eq:regulated_RWZ_equation} or eq.~\eqref{eq:RWeqChandra}), we find the recurrence relation at second order:
\begin{align}
    \alpha^{(2)}_0 a_{1}^{(2)} + \beta^{(2)}_0 a_0^{(2)} &= b_0 \; , \\
    \alpha^{(2)}_n a_{n+1}^{(2)} + \beta^{(2)}_n a_n^{(2)} + \gamma^{(2)}_n a_{n-1}^{(2)} &= b_n \; ,
\end{align}
where $\alpha_n^{(2)}$, $\beta_n^{(2)}$ and $\gamma_n^{(2)}$ take the same expression than at first order with the replacement $\rho_i \rightarrow \rho_1 + \rho_2$, and the coefficients $b_n$ are related to an expansion of the regularized source term,
\begin{equation}
    \frac{r^3}{r-2M} \mathfrak S(r) = M A_1(r) A_2(r) \sum_{n=0}^{\infty} b_{n} \bigg( 1 - \frac{2M}{r} \bigg)^n \; ,
\end{equation}
and similarly for $\bar{\mathfrak S}$. 
Notice that the good asymptotic behavior of the source term $\mathfrak S$ precisely means that the left-hand side of this equation is proportional to the product of amplitudes $A_1(r) A_2(r)$ asymptotically and that the series $\sum_n b_n$ converges. We can easily obtain the coefficients $b_n$ from a series expansion of the source in the variable $u=1-2M/r$. We then numerically solve the recurrence relation at second order in a way very similar to the first order case, the main difference being that the frequency $\omega = \omega_1 + \omega_2$ is imposed so that we instead have to find a value for $a_0^{(2)}$ such that the series $\sum_n a_n^{(2)}$ converges. Finally, the nonlinear ratio is easily found as the asymptotic limit of $\Psi / (\psi_{1}^{(1)} \psi_{2}^{(1)})$.

\subsection{Summing over parities} \label{sec:sum_over_parities}

In the QNM literature, linear modes amplitudes are often indexed by their overtones and mirror modes, but not by their parities. This is because of the well-known properties that the RW and Zerilli equations are isospectral~\cite{1975RSPSA.344..441C}, so that even and odd QNMs have the same frequencies. We thus can't read the amplitude of a  ``pure even mode'' or a ``pure odd mode'' from data, because there is no way to disentangle them when extracting the amplitudes of single-frequency components. 

Thus a quadratic mode amplitude is generated from the two full linear modes including both even and odd contributions in the waveform, reconstructed from the RW and Zerilli scalars through eq.~\eqref{eq:metric_reconstruction}. Schematically,
$\psi_1^{(1)} = \psi_{1,-}^{(1)} + \psi_{1,+}^{(1)}$ and $\psi_2^{(1)} = \psi_{2,-}^{(1)} + \psi_{2,+}^{(1)}$, although the complete expression of the metric should be read from~\cref{eq:spherical_decomposition1,eq:spherical_decomposition2,eq:spherical_decomposition3}. 
We would also like our formula for the source term to correctly describe the case where we look at a quadratic mode generated by the square of the \textit{same} linear mode $\psi_1^{(1)}$. In order to do this, we can keep our generic expression for the source where we assume that $\psi_1^{(1)}$ and $\psi_2^{(1)}$ are two different perturbations that linearly add in the waveform, and divide the source by a symmetry factor $\texttt{S}$ which is $\texttt{S}=1$ if $\psi_1^{(1)}$ and $\psi_2^{(1)}$ are different, and  $\texttt{S}=2$ if $\psi_1^{(1)}$ and $\psi_2^{(1)}$ are the same (i.e. $\ell_1=\ell_2$, $m_1=m_2$, $n_1=n_2$ and $\mathfrak m_1 = \mathfrak m_2$).

Taking into account these considerations, the regularized source term in~\eqref{eq:regulated_RWZ_equation} is now given by
\begin{equation}
    \label{eq:parity_sum_Zerilli}
    \mathfrak S(r) =  \frac{1}{\texttt{S}} \begin{cases}
      \mathfrak s_{--\rightarrow+} \psi_{1,-} \psi_{2,-} +  \mathfrak s_{++\rightarrow+} \psi_{1,+} \psi_{2,+}, & \text{if}\ \ell+\ell_1+\ell_2 \text{ even} \; , \\
    \mathfrak s_{+-\rightarrow+} \psi_{1,+} \psi_{2,-} + \mathfrak s_{-+\rightarrow+} \psi_{1,-} \psi_{2,+}, & \text{otherwise} \; .
    \end{cases}
\end{equation}
for the source term in the Zerilli equation, and
\begin{equation}
    \label{eq:parity_sum_RW}
    \mathfrak S(r) =  \frac{1}{\texttt{S}} \begin{cases}
      \mathfrak s_{--\rightarrow-} \psi_{1,-} \psi_{2,-} +  \mathfrak s_{++\rightarrow-} \psi_{1,+} \psi_{2,+}, & \text{if}\ \ell+\ell_1+\ell_2 \text{ odd} \; , \\
       \mathfrak s_{+-\rightarrow-}\psi_{1,+} \psi_{2,-} + \mathfrak s_{-+\rightarrow-} \psi_{1,-} \psi_{2,+}, & \text{otherwise} \; .
    \end{cases}
\end{equation}
for the source term in the RW equation.
To obtain these relations, we have used the parity selection rules explained in sec.~\ref{sec:linearVSQuad}, and denoted by e.g. $\mathfrak s_{+-\rightarrow+}$ the source term for the Zerilli equation in~\cref{eq:sourceSs} where the first mode $\omega_1$ has even parity and the second mode $\omega_2$ has odd parity. One should be careful that generically $\mathfrak s_{+-\rightarrow+} \neq \mathfrak s_{-+\rightarrow+}$ if the linear modes are not the same (our convention here is that the first index in $\mathfrak{s}_{p_1 p_2 \rightarrow p}$ refers to the first linear mode): one can be obtained from the other 
if we exchange \textit{all} the mode numbers of both linear modes $(1 \leftrightarrow 2)$, and similarly for the quadratic odd sector.
Thus, for a given $\ell+\ell_1+\ell_2$, we see that the amplitude of the corresponding quadratic mode (including both even and odd parity at second order) depends on four numbers which we can compute using the algorithm presented in sec.~\ref{sec:leaver}. 

In the previous literature on quadratic modes~\cite{Ma_2022,London_2014,Mitman:2022qdl, Cheung:2022rbm,Cheung:2023vki,Khera:2023oyf,Zhu:2024rej,Redondo-Yuste:2023seq,Lagos_2023,Nakano:2007cj,Ioka:2007ak,Bucciotti:2023ets,Perrone:2023jzq}, it has often been stated that nonlinearities can be encoded in a single number which is the ratio of quadratic to linear amplitudes. 
However, as this section makes clear, this ratio also depends on the parity content of the linear mode amplitudes, making it not a universal quantity that can be compared across different numerical simulations. This fact was recently highlighted in~\cite{Bourg:2024jme} to explain discrepancies across independent approaches in computing the ratio of amplitudes. Nevertheless, our present work shows that quadratic amplitudes depend on only four numbers multiplying the linear mode amplitudes, and these are the universal numbers that one should aim to measure in numerical relativity simulations. Therefore, quadratic amplitudes are still independent of the initial conditions in the sense that they are fixed (and given by our present work) once the initial conditions for the linear modes are fixed.

\section{Physical observables}
\label{sec:physical_observables}
Assuming the unique solution $\Psi^{(2)}$ of~\cref{eq:regulated_RWZ_equation} satisfying QNM boundary conditions is known, we are ready to reconstruct the quadratic metric perturbations. First in RW gauge (denoted with a tilde on metric perturbations), then we will transform it in the transverse traceless ($TT$) gauge (denoted with a superscript~$^{TT}$). Finally we will obtain the physical waveform plus and cross polarizations $\mathfrak h_+$ and $\mathfrak h_\times$. Conceptually, the logical chain is
\begin{equation}
    \Psi^{(2)}\Rightarrow\psi^{(2)}\Rightarrow\tilde h_{\mu\nu}\Rightarrow h_\pm^{TT}\Rightarrow \mathfrak h_+,\mathfrak h_\times \; .
\end{equation}
Since we will be interested in the asymptotic amplitude of the waveform, we will work in the asymptotic limit $r\rightarrow\infty$ whenever possible. The regulated master scalars in \cref{eq:regulated_master_scalar} were designed so that their asymptotic expansion at large $r$ takes the same form as at linear order (see \cref{eq:BC2nd}). We can then extract their amplitudes as~\footnote{Notice that our parameters $\mathcal{A}_+$ and $\mathcal{A}_-$ are denoted as $C^+$ and $C^-$ in the recent work~\cite{Bourg:2024jme}}
\begin{equation}
    \label{eq:asymptotic_psi_quadratic}
    \Psi_+^{(2)} \sim M \mathcal{A}_+^{(2)} e^{i \omega r_*},\qquad \Psi_-^{(2)} \sim M \mathcal{A}_-^{(2)} e^{i \omega r_*} \; ,
\end{equation}
the $M$ factor is added to have dimensionless amplitudes. At linear order we can work with the usual (unregulated) master scalars $\psi^{(1)}$
\begin{equation}
    \label{eq:asymptotic_psi_linear}
    \psi_+^{(1)} \sim M \mathcal{A}_+^{(1)} e^{i \omega r_*},\qquad \psi_-^{(1)} \sim M \mathcal{A}_-^{(1)} e^{i \omega r_*} \; .
\end{equation}
Linear and quadratic physical amplitudes are both needed because our main result will be their ratio.
To avoid clutter, we will write equal signs in place of $\sim$, even though we are dealing with asymptotic expansions. To evaluate derivatives in the asymptotic expansion, we recall that we will deal with functions of the form $r^k e^{i \omega(r_*-t)}$, where $k$ is an integer and $\omega$ the frequency of the mode (i.e. $\omega = \omega_i$ for linear modes and $\omega = \omega_1+\omega_2$ for quadratic modes): consequently $\partial_{r_*} \rightarrow i \omega$ and  $\partial_t \rightarrow - i \omega$ at leading order in the $1/r$ expansion.

\subsection{Reconstructing $\tilde h_{\mu\nu}$}
We split the discussion between linear and quadratic order. In both cases substituting $\psi^{(i)}$ into \cref{eq:metric_reconstruction} gives the metric in RW gauge. The main complication at quadratic order is that $\psi^{(2)}$ should be obtained from $\Psi^{(2)}$ using \cref{eq:regulated_master_scalar}.

\subsubsection{Linear modes}
Simple asymptotic expansions of \cref{eq:metric_reconstruction} at linear order give
\begin{align}
\label{eq:reconstruction_tildeh_linear}
     \tilde h_{tt}^{(1)} = \tilde h_{rr}^{(1)} = - \tilde h_{tr}^{(1)} = - \omega^2 r M \mathcal{A}_+^{(1)} e^{i \omega r_*} \; , \quad \tilde h_{t-}^{(1)} = - \tilde h_{r-}^{(1)} =  \frac{i \omega}{2} r M \mathcal{A}_-^{(1)} e^{i \omega r_*},\\
     \quad \tilde h_\circ^{(1)} = i \omega r^2 M \mathcal{A}_+^{(1)} e^{i \omega r_*} \; .
\end{align}

\subsubsection{Quadratic modes}
The necessary steps are just a matter of substitutions so we will refer the reader to the six notebooks \textquotedblleft Asymptotics\textquotedblright{} in the supplementary materials associated with this paper, where we give all formulas to reconstruct the quadratic metric components for any parity of the linear and quadratic perturbations.

To give an example, when an even and an odd linear perturbation couple to give an odd quadratic perturbation, the required $\Delta(r)$, defined in \cref{eq:delta_definition}, is
\begin{equation}
    \Delta_{+-\rightarrow -}(r) = -\mathcal{A}_{+,1}^{(1)}\mathcal{A}_{-,2}^{(1)}(-1)^m \frac{\omega_1^2}{\mu_\ell^2} \frac{\lambda_{1,\ell_2}}{\lambda_{1,\ell}} \sqrt{\frac{(2 \ell+1) (2 \ell_1+1) (2 \ell_2+1)}{4 \pi }} \begin{pmatrix}
        \ell_1&\ell_2&\ell\\
        0&1&-1
    \end{pmatrix}
    \begin{pmatrix}
        \ell_1&\ell_2&\ell\\
        m_1&m_2&-m
    \end{pmatrix}r \; ,
\end{equation}
where the parenthesis are $3j$ symbols and $\mathcal{A}_{+,i}$ and $\mathcal{A}_{-,i}$ are the asymptotic amplitudes of the RW and Zerilli variables defined in \cref{eq:asymptotic_psi_linear}. Then
\begin{equation}\label{eq:hrm2}
    \tilde h_{r-}^{(2)} =  -\frac{i \omega}{2} r M \mathcal{A}_-^{(2)} e^{i \omega r_*}+\frac{2r^2}{\mu^2_{\ell}}S_{r-}+i r \omega\frac{\Delta_{+-\rightarrow -}(r) \psi^{(1)}_+(r) \psi^{(1)}_-(r)}{2f(r)} \; ,
\end{equation}
where $S_{r-}$ goes to a constant at large $r$. As promised, the asymptotic behaviour of $\tilde h^{(2)}$ is identical to the linear order one because the divergent $\sim r^2$ term coming from the source cancels against our correcting term coming from $\Delta(r)$.

Before moving on, let us introduce a notation that will prove to be useful when relating RW gauge to TT gauge quantities:
\begin{align}
\label{eq:reconstruction_tildeh_quadratic}
     \tilde h_{tt}^{(2)} = \tilde h_{rr}^{(2)} = - \tilde h_{tr}^{(2)} = - \omega^2 r M \bar{\mathcal{A}}_+^{(2)} e^{i \omega r_*} \; , \quad \tilde h_{t-}^{(2)} = - \tilde h_{r-}^{(2)} =  \frac{i \omega}{2} r M \bar{\mathcal{A}}_-^{(2)} e^{i \omega r_*},\\
     \quad \tilde h_\circ^{(2)} = i \omega r^2 M \bar{\mathcal{A}}_+^{(2)} e^{i \omega r_*} \; .
\end{align}
where $\bar{\mathcal{A}}_+^{(2)}$ can be obtained from $\mathcal{A}_+^{(2)}$ and all terms quadratic in the linear amplitudes present in the source term (e.g. in eq.~\eqref{eq:hrm2}): explicit expressions are provided in the supplementary files.

\subsection{Reconstructing $h_+^{TT}$, $h_-^{TT}$}
We now need the expression for the asymptotic waveform in TT gauge, defined by the condition $h_{ab}^{TT} = h_{a \pm}^{TT} = h_\circ^{TT} = 0$ at leading order for large $r$, while only $h_\pm^{TT}$ are non-zero~\footnote{A more precise statement is $h_{ab}^{TT} = \mathcal{O}(r^{-2})$, $h_{a\pm}^{TT} = \mathcal{O}(r^{-1})$, $h_{\circ}^{TT} = \mathcal{O}(r^{0})$ and $h_{\pm}^{TT} = \mathcal{O}(r)$}. On the other hand, up to now we have worked in RW gauge where $\tilde h_{t+} = \tilde h_{r+} = \tilde h_+ = \tilde h_- = 0 $. We thus perform an infinitesimal diffeomorphism $\tilde h_{\mu \nu}^{(i)} \rightarrow \tilde h_{\mu \nu}^{(i)} + \Delta h_{\mu \nu}^{(i)}$ up to second order included, where
\begin{align}
    \Delta h_{\mu \nu}^{(1)} &=  \mathcal{L}_{\xi^{(1)}} \bar g_{\mu \nu} \nonumber \; , \\
    \Delta h_{\mu \nu}^{(2)} &= \mathcal{L}_{\xi^{(2)}} \bar g_{\mu \nu} + \frac{1}{2} \mathcal{L}^2_{\xi^{(1)}} \bar g_{\mu \nu} + \mathcal{L}_{\xi^{(1)}} \tilde h^{(1)}_{\mu \nu} \;.
    \label{eq:diffeo_on_h_RW_to_TT}
\end{align}
$\xi^{(i)}_\mu = (\zeta^{(i)}_a, Z^{(i)}_A)$ is a vector to be determined from the TT gauge condition. We decompose the gauge transformation~\eqref{eq:diffeo_on_h_RW_to_TT} in tensor spherical harmonics, splitting as before the discussion between linear and quadratic modes.

\subsubsection{Linear modes} \label{sec:RW_to_TT_linear}

In this case, the components of the gauge transformation we are interested in are given in~\cite{Spiers:2023mor}
\begin{align}
    \Delta h_{a+}^{(1)} &= \zeta_a^{(1)} + r^2 e^{i \omega t} \partial_a \big[ e^{-i \omega t} Z_+^{(1)} \big] \; ,\\
    \Delta h_{a-}^{(1)} &= r^2 e^{i \omega t} \partial_a \big[ e^{-i \omega t} Z_-^{(1)} \big] \; , \\
    \Delta h_{\circ}^{(1)} &= 2 r f(r) \zeta_r^{(1)} - \ell(\ell+1) r^2 Z_+^{(1)} \; , \\
    \Delta h_{\pm}^{(1)} &= 2 r^2 Z_\pm^{(1)} \; ,
\end{align}
where $Z_A^{(i)} = \sum_{\ell m \omega} e^{-i \omega t} \big[ Z_+^{(i)} Y_A^{\ell m} + Z_-^{(i)} X_A^{\ell m} \big]$ has been decomposed in vector harmonics following the definitions in app.~\ref{app:tensor_harmonics}.
Imposing TT gauge and working as before in the asymptotic limit, we find using~\cref{eq:reconstruction_tildeh_linear}
\begin{equation}
    Z_\pm^{(1)} = \frac{M \mathcal{A}_\pm^{(1)}}{2r} e^{i \omega r_*} \; , \quad \zeta_t^{(1)} = - \zeta_r^{(1)} = \frac{i \omega r}{2} M \mathcal{A}_+^{(1)} e^{i \omega r_*} \; ,
\end{equation}
so that the nonzero components of the metric in TT gauge are
\begin{equation}
    h_\pm^{(1),TT} = r M \mathcal{A}_\pm^{(1)} e^{i \omega r_*} \; .
\end{equation}

\subsubsection{Quadratic modes}

At second order, the components of the gauge transformation~\eqref{eq:diffeo_on_h_RW_to_TT} are very similar to the first-order ones, except for the presence of a supplementary tensor $H_{\mu \nu} := \mathcal{L}^2_{\xi^{(1)}} \bar g_{\mu \nu} /2 + \mathcal{L}_{\xi^{(1)}} \tilde h^{(1)}_{\mu \nu}$, which is quadratic in the linear perturbations generating the nonlinear mode. When we expand $\xi^{(1)}$ and $\tilde h_{\mu \nu}^{(1)}$ in a spherical harmonic basis, $H_{\mu \nu}$ will consist of a sum of product of first-order modes, of which we select the nonlinear mode corresponding to a frequency $\omega = \omega_1 + \omega_2$. As in sec.~\ref{sec:sum_over_parities}, we include a symmetry factor $\texttt{S}$ to also cover the case where the quadratic mode is generated from the product of the same linear mode (see also app.~\ref{app:toy_model}). Computing the components of $H_{\mu \nu}$ in a tensor spherical harmonic basis in the limit $r \rightarrow \infty$ is tedious but straightforward using the methodology sketched in app.~\ref{app:angular_integrals}. We provide their complete expression in the companion \textsc{Mathematica} notebooks, while reporting here only one component for the sake of example:
\begin{align}
\begin{split}
     H_{tt} = H_{rr} = - H_{tr} &= i (-1)^{m+1} \frac{ M^2 r e^{i\omega r_*}}{4 \texttt{S}} \mathcal{A}_{+,1}^{(1)} \mathcal{A}_{+,2}^{(1)} \\
     &\times C_{\ell_1 m_1 0, \ell_2 m_2 0, \ell (-m) 0} \big(\omega_1^3 + \omega_1^2 \omega_2 + \omega_1 \omega_2^2 + \omega_2^3 \big) \; ,
    \label{eq:Htt}
\end{split}
\end{align}
where the coefficient $C_{\ell_1 m_1 0, \ell_2 m_2 0, \ell m 0}$ is given in~\cref{eq:generalized_CB_coefficients_value}. A nontrivial result we obtain is that the components of $H_{\mu \nu}$ are $\propto r$ (except $H_\circ \propto r^2$), so that they have the same scaling at large $r$ as the components of $\tilde h_{\mu \nu}$ at linear order.

We can now follow the same steps as in sec.~\ref{sec:RW_to_TT_linear} to find the diffeomorphism $\xi_\mu^{(2)}$ to go from RW to TT gauge at second order:
\begin{align}
    \begin{split}
        Z_-^{(2)} &=  \frac{M \bar{\mathcal{A}}_-^{(2)}}{2r} e^{i \omega r_*} + \frac{H_{t-}}{i \omega r^2} \; , \\
        Z_+^{(2)} &=  \frac{M \bar{\mathcal{A}}_+^{(2)}}{2r} e^{i \omega r_*} + \frac{H_{\circ}}{2 i \omega r^3} + \frac{H_{t+}}{i \omega r^2} \; , \\
        \quad \zeta_t^{(2)} &= - \zeta_r^{(2)} = \frac{i \omega r}{2} M \bar{\mathcal{A}}_+^{(2)} e^{i \omega r_*} + \frac{H_\circ}{2r} \; .
    \end{split}
\end{align}
Thus, the nontrivial metric components in TT gauge are
\begin{align}
    \begin{split}
        h_+^{(2),TT} &= r M \bar{\mathcal{A}}_+^{(2)} e^{i \omega r_*} + \frac{H_\circ}{i \omega r} + 2 \frac{H_{t+}}{i \omega} + H_+ \; , \\
        h_-^{(2),TT} &= r M \bar{\mathcal{A}}_-^{(2)} e^{i \omega r_*} + 2 \frac{H_{t-}}{i \omega} + H_- \; .
    \end{split}
\end{align}
Once again, in the following, we will introduce a convenient notation:
\begin{equation}\label{eq:hpmTT}
    h_\pm^{(i), TT} = M r \tilde{\mathcal{A}}_\pm^{(i)} e^{i \omega r_*} \; ,
\end{equation}
where $\tilde{\mathcal{A}}_\pm^{(1)} = \mathcal{A}_\pm^{(1)}$ at first order, while at second order $\tilde{\mathcal{A}}_\pm^{(2)}$ includes the contributions of the various $H_{\mu \nu}$ components (quadratic in the linear amplitudes) into the final second-order amplitude in TT gauge.

\subsection{Reconstructing $\mathfrak h_+$, $\mathfrak h_\times$}
We are now ready to connect our computation with physical predictions. To begin with, we will define a spacetime tetrad $e_a^\mu$ as
\begin{equation}
    e_0^\mu = f(r)^{-1/2} \delta_0^\mu \; , \quad e_r^\mu = f(r)^{1/2} \delta_r^\mu \; , \quad e_\theta^\mu = r^{-1} \delta_\theta^\mu \ ; , \quad e_\phi^\mu = (r \sin \theta)^{-1} \delta_\phi^\mu \; ,
\end{equation}
where we recall that $f(r) = 1-2M/r$. The physical waveforms of the $+$ and $\times$ polarizations are encoded by the variables $\mathfrak h_+$ and $\mathfrak h_\times$ defined as~\cite{Maggiore:1900zz}
\begin{equation}
    \mathfrak h_+ = \frac{1}{2} h_{\mu \nu}^{TT} (e_+)^{\mu \nu} \; , \quad \mathfrak h_\times = \frac{1}{2} h_{\mu \nu}^{TT} (e_\times)^{\mu \nu} \; ,
\end{equation}
where the polarization tensors are given by
\begin{equation} \label{eq:polarizationTensors}
    (e_+)^{\mu \nu} = e_\theta^\mu e_\theta^\nu - e_\phi^\mu e_\phi^\nu \; , \quad (e_\times)^{\mu \nu} = e_\theta^\mu e_\phi^\nu + e_\phi^\mu e_\theta^\nu \; .
\end{equation}
Explicitly, we have
\begin{equation} \label{eq:defhphc}
    \mathfrak h_+ = \frac{1}{2r^2} \bigg( h_{\theta \theta}^{TT} - \frac{h^{TT}_{\phi \phi}}{\sin^2\theta} \bigg) \; , \quad \mathfrak h_\times = \frac{h_{\theta \phi}^{TT}}{r^2 \sin \theta} \; .
\end{equation}
Using the full expression of the metric in eq.~\eqref{eq:spherical_decomposition3}, the expression of the even and odd part of the metric in~\eqref{eq:hpmTT} and identities for the spin-weighted spherical harmonics (see app.~\ref{app:tensor_harmonics}, in particular~\cref{eq:YAB_from_2Y,eq:XAB_from_2Y}), we can get a compact expression for the complex strain, which is the most commonly used quantity in the NR literature:
\begin{equation}\label{eq:NPscalar}
    \mathfrak h_+ - i \mathfrak h_\times = \frac{M}{r} \sum_{\ell m \mathcal{N}}  \mathcal{A}_{\ell m \mathcal{N}} e^{i \omega_{\ell \mathcal{N}} (r_*-t)} \hphantom{a}_{-2}Y^{\ell m}(\theta, \phi) \; .
\end{equation}
In this equation, we have reintroduced explicitly the sum over all modes (both linear and quadratic) needed to recover the full waveform from individual modes amplitudes; thus $\mathcal{N}$ is the supplementary mode number, i.e. $\mathcal{N} = (n, \mathfrak m)$ for linear modes and $\mathcal{N} = (\ell_1,m_1,n_1,\mathfrak m_1) \times (\ell_2,m_2,n_2,\mathfrak m_2)$ for nonlinear modes (see section~\ref{sec:linearVSQuad}). This provides us with a way to read the physical amplitude of the mode $ \mathcal{A}_{\ell m \mathcal{N}}$ defined  from the individual parities amplitudes as follows
\begin{equation}\label{eq:Almn}
    \mathcal{A}_{\ell m \mathcal{N}} := \frac{\lambda_{2,\ell}}{2} \big( \tilde{\mathcal{A}}_{+, \ell m \mathcal{N}} - i \tilde{\mathcal{A}}_{-, \ell m \mathcal{N}} \big) \; ,
\end{equation}
where this relation is valid both at first and second order (so that we do not write explicitly the perturbation order), and we recall that $\lambda_{2,\ell} = \sqrt{\ell(\ell+1)(\ell+2)(\ell-1)}$.
On the other hand, we would also like to be able to recover the individual parities amplitudes $\tilde{\mathcal{A}}_{+, \ell m \mathcal{N}}$ and $\tilde{\mathcal{A}}_{-, \ell m \mathcal{N}}$ from the physical amplitude $\mathcal{A}_{\ell m \mathcal{N}}$. To do this, we can remark that $\mathfrak h_+$ and $\mathfrak h_\times$ are real to obtain
\begin{equation}\label{eq:hp+ihtimes1}
    \mathfrak h_+ + i \mathfrak h_\times = \frac{M}{r} \sum_{\ell m \mathcal{N}} (-1)^m  \mathcal{A}_{\ell m \mathcal{N}}^* e^{-i \omega^*_{\ell \mathcal{N}} (r_*-t)} \hphantom{a}_{2}Y^{\ell (-m)}(\theta, \phi) \; ,
\end{equation}
where we have used the relation $\vphantom{|}_{-2}Y^{\ell m *}(\theta, \phi) = (-1)^m \vphantom{|}_{2}Y^{\ell (-m) }(\theta, \phi)$. Now, we can use the fact that mirror modes always come in pair with standard modes to relabel the sum indices. Let us denote by $\bar{\mathcal{N}}$ the same mode number than $\mathcal{N}$ but in which we transformed regular modes to mirror modes and vice-versa (i.e., we flipped the sign of the real part of $\omega$) and we flipped the sign of $m$: $\bar{\mathcal{N}} = (n, - \mathfrak m)$ for linear modes and $\bar{\mathcal{N}} = (\ell_1,-m_1,n_1,-\mathfrak m_1) \times (\ell_2,-m_2,n_2,-\mathfrak m_2)$ for nonlinear modes. By relabelling the summation indices in~\eqref{eq:hp+ihtimes1} and using the relation $\omega_{\ell \bar{\mathcal{N}}}^* = - \omega_{\ell \mathcal{N}}$ we get
\begin{equation}\label{eq:hp+ihtimes2}
    \mathfrak h_+ + i \mathfrak h_\times = \frac{M}{r} \sum_{\ell m \mathcal{N}} (-1)^m  \mathcal{A}_{\ell (-m) \bar{\mathcal{N}}}^* e^{i \omega_{\ell \mathcal{N}} (r_*-t)} \hphantom{a}_{2}Y^{\ell m}(\theta, \phi) \; .
\end{equation}
On the other hand, we can follow the same procedure as explained above eq.~\eqref{eq:NPscalar} to get $\mathfrak h_+ + i \mathfrak h_\times$ from the individual parities amplitudes, which lead to
\begin{equation}\label{eq:AlmnMirror}
     (-1)^m \mathcal{A}_{\ell (-m) \bar{\mathcal{N}}}^* = \frac{\lambda_{2,\ell}}{2} \big( \tilde{\mathcal{A}}_{+, \ell m \mathcal{N}} + i \tilde{\mathcal{A}}_{-, \ell m \mathcal{N}} \big) \; .
\end{equation}
By inverting the system~\eqref{eq:Almn}-\eqref{eq:AlmnMirror} we can recover the individual parities amplitudes from the strain amplitudes as
\begin{align} \label{eq:AEfromA}
    \tilde{\mathcal{A}}_{+, \ell m \mathcal{N}} &= \frac{1}{\lambda_{2,\ell}} \big(     \mathcal{A}_{\ell m \mathcal{N}} + (-1)^m \mathcal{A}_{\ell (-m) \bar{\mathcal{N}}}^* \big) \; , \\
    \tilde{\mathcal{A}}_{-, \ell m \mathcal{N}} &= \frac{i}{\lambda_{2,\ell}} \big(     \mathcal{A}_{\ell m \mathcal{N}} - (-1)^m \mathcal{A}_{\ell (-m) \bar{\mathcal{N}}}^* \big) \label{eq:AofromA} \; .
\end{align}
Thus, we see that the amplitude of a regular mode for a given parity depends on both the amplitudes of regular and mirror modes in the strain. 
We can now follow our main algorithm for computing the amplitudes of quadratic modes, which is the following. First, use eqs.~\eqref{eq:AEfromA}-\eqref{eq:AofromA} to read the individual parities amplitudes of linear modes from the strain. Then, use our Leaver code to compute the individual parities amplitudes of the corresponding quadratic mode. Finally, use eq.~\eqref{eq:Almn} to find the amplitude of the quadratic mode in the strain. 

Our results at this point are rather generic. In the next section we will study how reflection (or equatorial) symmetry will allow us to simplify some of our formulas while giving us insight into how quadratic modes amplitudes depend on the linear ones.

\subsection{Reflection symmetry}

The results of the previous section can be simplified if we impose a symmetry which is commonly assumed in the QNM literature~\cite{Berti:2007fi, Berti:2007zu, Isi:2021iql, Yi:2024elj}, namely reflection (or equatorial) symmetry. Indeed, if the underlying cause for BH perturbations is a BBH merger, and if we neglect the spins of the progenitor BHs as a first approximation, then the system is reflection symmetric with respect to the plane of motion of the BBH. We prove in Appendix~\ref{app:reflection_symmetry} that it imposes the following relation between the amplitude of regular and mirror modes:
\begin{align} \label{eq:mirrorModeAmplitude}
    \mathcal{A}_{\ell (-m) \bar{\mathcal{N}}}^* &= (-1)^\ell \mathcal{A}_{\ell m \mathcal{N}} \; .
\end{align}
Waveforms computed assuming reflection symmetry have a circular polarization pattern up to angular factors, see e.g.~\cite{Isi:2021iql}. Let us now introduce a \textit{polarization parameter} $\kappa_{\ell m \mathcal{N}}$ encoding deviations from equatorial symmetry:
\begin{equation}
    \kappa_{\ell m \mathcal{N}} = (-1)^m \frac{\mathcal{A}_{\ell (-m) \bar{\mathcal{N}}}^*}{\mathcal{A}_{\ell m \mathcal{N}}} \; .
\end{equation}
We see that $\kappa_{\ell m \mathcal{N}}=(-1)^{\ell+m}$ if we assume reflection symmetry, however we will keep $\kappa_{\ell m \mathcal{N}}$ arbitrary in our final results. Using \cref{eq:AEfromA,eq:AofromA} there is now a one-to-one correspondence between the individual parities amplitudes $ \tilde{\mathcal{A}}_{+, \ell m \mathcal{N}}$, $ \tilde{\mathcal{A}}_{-, \ell m \mathcal{N}}$ and the strain amplitude together with the polarization parameter, given by
\begin{align} \label{eq:AEfromAandKappa}
    \tilde{\mathcal{A}}_{+, \ell m \mathcal{N}} &= \frac{\mathcal{A}_{\ell m \mathcal{N}}}{\lambda_{2,\ell}} \big(     1 + \kappa_{\ell m \mathcal{N}} \big) \; , \\
    \tilde{\mathcal{A}}_{-, \ell m \mathcal{N}} &= i \frac{\mathcal{A}_{\ell m \mathcal{N}}}{\lambda_{2,\ell}} \big(     1 -  \kappa_{\ell m \mathcal{N}} \big) \label{eq:AofromAandKappa} \; ,
\end{align}
and the inverse relations read
\begin{align} \label{eq:AfromAoAe}
    \mathcal{A}_{\ell m \mathcal{N}} &= \frac{\lambda_{2,\ell}}{2} \big( \tilde{\mathcal{A}}_{+, \ell m \mathcal{N}} - i \tilde{\mathcal{A}}_{-, \ell m \mathcal{N}} \big) \; , \\
    \kappa_{\ell m \mathcal{N}} &=  \frac{\tilde{\mathcal{A}}_{+, \ell m \mathcal{N}} + i \tilde{\mathcal{A}}_{-, \ell m \mathcal{N}}}{\tilde{\mathcal{A}}_{+, \ell m \mathcal{N}} - i \tilde{\mathcal{A}}_{-, \ell m \mathcal{N}}} \label{eq:kappaFromAoAe} \; .
\end{align}
Here we can notice that, if one assumes reflection symmetry $\kappa_{\ell m \mathcal{N}}=(-1)^{\ell+m}$, then the parity of the mode is completely given by the parity of $\ell + m$, i.e. $h_-$ vanishes for $\ell+m$ even and vice-versa. 

\section{Final ratio of amplitudes and its symmetries} \label{sec:final_ratio}

We are now ready to assemble all of our formulas together to obtain the ratio of quadratic amplitudes to linear ones. We will express our results in terms of the amplitudes and polarization parameters entering the strain in~\cref{eq:NPscalar}, because these are the most relevant for comparing our results with NR simulations. 
Taking into account the sum over parities mentioned in~\cref{sec:sum_over_parities} and all the previous results obtained in this section (in particular~\cref{eq:AfromAoAe,eq:kappaFromAoAe}), we can express the quadratic mode amplitude and polarization parameter in the following way. 

If $\ell + \ell_1 + \ell_2$ is even:
\begin{align} \label{eq:quadModeAmplEven}
    \frac{\mathcal{A}_{\ell m \mathcal{N}}^{(2)}}{\mathcal{A}_{1}^{(1)} \mathcal{A}_{2}^{(1)}} &= \frac{1}{4} \big[ \alpha_+ + \beta_+ \big] \; , \quad \kappa^{(2)}_{\ell m \mathcal{N}} = \frac{\alpha_+-\beta_+}{\alpha_+ +\beta_+} \; , \\
    \alpha_+ &= \mathcal{R}_{--\rightarrow+} \big( 1- \kappa_1^{(1)} \big) \big( 1- \kappa_2^{(1)} \big) + \mathcal{R}_{++\rightarrow+} \big( 1+ \kappa_1^{(1)} \big) \big( 1+ \kappa_2^{(1)} \big) \nonumber \; , \\
    \beta_+ &= \mathcal{R}_{+-\rightarrow-} \big( 1+ \kappa_1^{(1)} \big) \big( 1- \kappa_2^{(1)} \big) + \mathcal{R}_{-+\rightarrow-}  \big( 1- \kappa_1^{(1)} \big) \big( 1+ \kappa_2^{(1)} \big) \nonumber \; ,
\end{align}
where to shorten the notation we have denoted $\mathcal{A}_i^{(1)} := \mathcal{A}_{\ell_i m_i \mathcal{N}_i}^{(1)}$, $\kappa_i^{(1)} := \kappa_{\ell_i m_i \mathcal{N}_i}^{(1)}$.

If  $\ell + \ell_1 + \ell_2$ is odd:
\begin{align} \label{eq:quadModeAmplOdd}
    \frac{\mathcal{A}_{\ell m \mathcal{N}}^{(2)}}{\mathcal{A}_{1}^{(1)} \mathcal{A}_{2}^{(1)}} &= \frac{1}{4} \big[ \alpha_- + \beta_- \big] \; , \quad \kappa^{(2)}_{\ell m \mathcal{N}} = \frac{\alpha_- -\beta_-}{\alpha_- +\beta_-} \; , \\
    \alpha_- &= \mathcal{R}_{+-\rightarrow+} \big( 1+ \kappa_1^{(1)} \big) \big( 1- \kappa_2^{(1)} \big) + \mathcal{R}_{-+\rightarrow+} \big( 1- \kappa_1^{(1)} \big) \big( 1+ \kappa_2^{(1)} \big) \nonumber \; ,  \\
    \beta_- &= \mathcal{R}_{--\rightarrow-} \big( 1- \kappa_1^{(1)} \big) \big( 1- \kappa_2^{(1)} \big) + \mathcal{R}_{++\rightarrow-} \big( 1+ \kappa_1^{(1)} \big) \big( 1+ \kappa_2^{(1)} \big) \nonumber \; .
\end{align}
In this equation, the ratio of amplitudes for the individual parities $\mathcal{R}_{p_1 p_2 \rightarrow p}$ are computed with our Leaver code, so that~\cref{eq:quadModeAmplEven,eq:quadModeAmplOdd} completely fix the quadratic mode amplitudes and polarizations as a function of the linear mode amplitudes and polarizations. It is easy to notice that assuming $\kappa_i^{(1)} = \pm 1$ implies $\kappa^{(2)}=\pm 1$ in all cases, so that if one imposes equatorial symmetry on the linear modes amplitudes then quadratic amplitudes will automatically respect the same symmetry. 
As already discussed in section~\ref{sec:sum_over_parities}, we observe that for a fixed parity of $\ell+\ell_1+\ell_2$, the amplitudes of quadratic modes depend on four numbers (in addition to the initial conditions of the linear mode), which we compute with our code. 

In practise, we need only to compute the ratios $\mathcal{R}_{p_1 p_2 \rightarrow p}$ for regular modes at second order (i.e. modes with positive real part of $\omega_1+\omega_2$). This is because we can use the following relations in order to compute second-order mirror modes amplitudes and polarizations:
\begin{equation}
    \mathcal{A}_{\ell (-m) \bar{\mathcal{N}}} = (-1)^m \kappa_{\ell m \mathcal{N}}^* \mathcal{A}_{\ell m \mathcal{N}}^* \; , \quad \kappa_{\ell (-m) \bar{\mathcal{N}}} = \frac{1}{\kappa_{\ell m \mathcal{N}}^*} \; .
\end{equation}
Thus, at second order the ratio of amplitudes for mirror modes reads
\begin{equation}
     \frac{\mathcal{A}_{\ell (-m) \bar{\mathcal{N}}}^{(2)}}{\mathcal{A}_{\ell_1 (-m_1) \bar{\mathcal{N}}_1}^{(1)} \mathcal{A}_{\ell_2 (-m_2) \bar{\mathcal{N}}_2}^{(1)} } =  \bigg( \frac{\kappa_{\ell m \mathcal{N}}^{(2)}}{\kappa_1^{(1)} \kappa_2^{(1)}} \frac{\mathcal{A}_{\ell m \mathcal{N}}^{(2)}}{\mathcal{A}_{1}^{(1)} \mathcal{A}_{2}^{(1)}} \bigg)^* \; .
\end{equation}

We provide our results in companion files~\cite{csvQuadratic}, where we give all the values of the ratios $\mathcal{R}_{p_1 p_2 \rightarrow p}$ entering~\cref{eq:quadModeAmplEven,eq:quadModeAmplOdd} for $2 \leq \ell, \ell_1, \ell_2 \leq 7$ respecting the Clebsch-Gordan rules $|\ell_1-\ell_2|\leq \ell \leq \ell_1+\ell_2$. We also include the cases where one of the linear amplitudes is an overtone $n=1$ or a mirror mode $\mathfrak m = -$. 
We furthermore provide a convenience Python function to directly compute the quadratic ratios $ \mathcal{A}_{\ell m \mathcal{N}}^{(2)}/\mathcal{A}_{1}^{(1)} \mathcal{A}_{2}^{(1)}$ either assuming reflection symmetry or supplementing $\kappa_1^{(1)}$, $\kappa_2^{(1)}$.

\subsection{Explicit values of the ratio}

If we assume reflection symmetry, then $\kappa_i = (-1)^{\ell_i+m_i}$ and the ratio of amplitudes is a single number that we compute with our code. In our companion paper~\cite{Bucciotti:2024zyp} we have plotted this ratio for several parent linear modes; while referring the reader to this paper for more details, let us just highlight here that a combination of a regular mode with a mirror mode can give appreciable nonlinear ratios in the $\ell=2$ multipole, which have not yet been observed in NR simulations. For example, we find
\begin{equation}
    \frac{\mathcal{A}^{(2)}_{21(220+)\times(2(-1)0-)}}{\mathcal{A}_{220+}^{(1)} \mathcal{A}_{2(-1)0-}^{(1)}} \simeq 0.104 e^{-1.59i} \; \text{(reflection symmetry).}
\end{equation}
On the other hand, we find for the loudest quadratic mode the following ratio of amplitudes:
\begin{equation}
    \frac{\mathcal{A}^{(2)}_{44(220+)\times(220+)}}{\big( \mathcal{A}_{220+}^{(1)}\big)^2} \simeq 0.154 e^{-0.068i} \; \text{(reflection symmetry),}
\end{equation}
in good agreement with NR simulations~\cite{London_2014,Ma_2022,Mitman:2022qdl,Cheung:2022rbm,Cheung:2023vki,Zhu:2024rej,Redondo-Yuste:2023seq}. 

As stated before, reflection symmetry is approximately satisfied in most NR simulations of mergers~\cite{Berti:2007fi}; however the symmetry is broken by the spins of the progenitor BHs and there will be small deviations from it in actual ringdown waveforms. Our results show that in this case the ratio of amplitudes depends on the linear modes initial conditions through their polarization parameters $\kappa_i^{(1)}$; this dependence could explain the discrepancies in the literature when computing the ratio of amplitudes in different physical settings~\cite{Ma:2024qcv,Zhu:2024rej,Redondo-Yuste:2023seq,Cheung:2023vki}. The dependence of the nonlinear ratio on linear modes polarizations (equivalently, on their polarity) has been analyzed in the recent~\cite{Bourg:2024jme};  we have verified that our generic formula~\eqref{eq:quadModeAmplEven} agrees with their figure 1 for the value of the dominant nonlinear mode as a function of $\kappa_i^{(1)}$. We refer the reader to that article for more details on this point.

\subsection{Two selection rules for the quadratic modes} \label{sec:theorem}

There is an additional symmetry in quadratic modes that is worth mentioning. To begin with, we notice that the $m_1,m_2,m$ dependence of the source (and hence of the ratio of amplitudes) is completely captured by a $3j$ symbol. This agrees with expectations based on the Wigner-Eckart theorem that rotational symmetry fixes the source for all $m_1,m_2,m$ if we know it for a single configuration, provided its $3j$ symbol is nontrivial. More importantly, a $3j$ symbol has the following symmetry property:
\begin{equation} \label{eq:symmetry3j}
    \begin{pmatrix}
        \ell_1&\ell_2&\ell\\
        m_1&m_2&-m
    \end{pmatrix} = (-1)^{\ell_1+\ell_2+\ell} \begin{pmatrix}
        \ell_2&\ell_1&\ell\\
        m_2&m_1&-m
    \end{pmatrix} \; .
\end{equation}
Thus, if $\ell_1+\ell_2+\ell$ is odd and $\ell_1=\ell_2$, $m_1=m_2$, the $3j$ symbol vanishes which implies a first selection rule:
\paragraph{Rule 1.}
    {\it Quadratic modes vanish if $\ell_1=\ell_2$, $m_1=m_2$ and $\ell$ is odd.}
\smallskip

Notice that it is valid regardless of the value of $n_1$, $n_2$, $\mathfrak m_1$, $\mathfrak m_2$, $\kappa_1^{(1)}$ and $\kappa_2^{(1)}$, and that we are focussing on the vanishing of the strain amplitude $\mathcal{A}^{(2)}_{\ell m \mathcal{N}}$ where all parities are summed over as in eq.~\eqref{eq:quadModeAmplOdd}. 
However, the antisymmetry~\eqref{eq:symmetry3j} of the $3j$ symbol (for $\ell+\ell_1+\ell_2$ odd) implies another, less straightforward, selection rule for quadratic modes. To see this, let us define a \textit{normalized} quadratic ratio $\hat{\mathcal R}$ by dividing $\mathcal{R}$ by its factoring $3j$ symbol. From the previous consideration on the Wigner-Eckart theorem, we get the property that $\hat{\mathcal R}$ does not depend on $m$, $m_1$, $m_2$. Furthermore, the antisymmetry~\eqref{eq:symmetry3j} together with the symmetry of the source term by exchange of \emph{all} the quantum numbers of the linear modes $(1 \leftrightarrow 2)$
implies that the normalized ratio is antisymmetric under $(1 \leftrightarrow 2)$ if $\ell + \ell_1 + \ell_2$ is odd. 

This means that $\hat{\mathcal R}$ vanishes if all quantum numbers of modes 1 and 2 are the same \textit{except} for $m_1$ and $m_2$, which can be arbitrary. Thus the ratios $\hat{\mathcal{R}}_{--\rightarrow-}$ and $\hat{\mathcal{R}}_{++\rightarrow-}$ in~\eqref{eq:quadModeAmplOdd} vanish in this case, but $\hat{\mathcal{R}}_{+-\rightarrow-}$ does not necessarily vanish since $p_1 \neq p_2$. However, we still get that $\hat{\mathcal{R}}_{-+\rightarrow-} = - \hat{\mathcal{R}}_{+-\rightarrow-}$ under $(1 \leftrightarrow 2)$. Thus the strain amplitude in eq.~\eqref{eq:quadModeAmplOdd} will vanish if $(1+\kappa_1^{(1)})(1-\kappa_2^{(1)}) = (1-\kappa_1^{(1)})(1+\kappa_2^{(1)})$, i.e. if $\kappa_1^{(1)}=\kappa_2^{(1)}$.
These considerations lead us to the following second selection rule:
\paragraph{Rule 2.}
    {\it Quadratic modes vanish if $\ell_1=\ell_2$, $n_1=n_2$, $\mathfrak m_1 = \mathfrak m_2$, $\kappa_1^{(1)}=\kappa_2^{(1)}$ and $\ell$ is odd}
\smallskip

Notice that this time the rule is valid regardless of the values of $m_1$, $m_2$.
Furthermore, if we assume reflection symmetry, $\kappa_1^{(1)}=\kappa_2^{(1)}$ means that $\ell_1+m_1$ has to be of the same parity than $\ell_2+m_2$.
This explains the vanishing of some of the quadratic modes plotted in Figure 1 of our companion article~\cite{Bucciotti:2024zyp}.

\section{Conclusions}
\label{sec:conclusions}

In this article we have developed the tools for an accurate computation of the amplitudes and polarization parameters of all quadratic QNMs for a \sch BH. We give in a companion package~\cite{csvQuadratic} the result of our Leaver code for quadratic modes with $2 \leq \ell, \ell_1, \ell_2 \leq 7$, which should be enough for nonlinear ringdown analysis with data from next-generation interferometers. Our work paves the way towards improved ringdown modelling as well as accurate tests of GR.
Indeed, one expects that in modified theories of gravity where BH possess scalar hair~\cite{herdeiro2015asymptotically} or a superradiant scalar cloud~\cite{Bertone:2019irm}, the ratio of quadratic to linear amplitudes would be different in data compared to our vacuum GR computation. This opens up a possibility for a new test of GR in the nonlinear regime.

Our final result in section~\ref{sec:final_ratio} shows that quadratic QNM amplitudes and polarizations are completely fixed once linear amplitudes are known, however the ratio of quadratic to linear amplitudes itself depends on the parity content of linear modes. In this sense, the ratio of amplitudes is not a universal quantity that should be the same across different initial conditions, as was emphasised in the recent work~\cite{Bourg:2024jme}. In the most generic case, quadratic modes amplitudes depend on four numbers that can be considered (like linear modes frequencies) to be fundamental properties of GR, related to the four different ways to combine two linear modes with different parities. We also proved two selection rules implying the vanishing of quadratic modes in some particular cases.
Although the dependence of quadratic modes on spin seems to be rather weak from Numerical Relativity simulations~\cite{Cheung:2023vki}, it still remains important to extend our results to the Kerr case, where the ratio of amplitudes will be promoted to a function of the spin of the remnant BH. Although technically challenging, third-order perturbation theory would also be an interesting avenue for future work. At this order the changing mass of the BH background begins to manifest itself in the ringdown signal and this effect may be relevant in analyzing the ringdown in black hole mergers~\cite{May:2024rrg}.
We plan to address these issues in the near future.

\acknowledgments
A. Kuntz acknowledges support from the European Union's H2020 ERC Consolidator Grant ``GRavity from Astrophysical to Microscopic Scales'' (Grant No. GRAMS-815673), the PRIN 2022 grant ``GUVIRP - Gravity tests in the UltraViolet and InfraRed with Pulsar timing'', and the EU Horizon 2020 Research and Innovation Programme under the Marie Sklodowska-Curie Grant Agreement No. 101007855. This project made use of the Black Hole Perturbation Toolkit~\cite{BHPToolkit}. We would like to thank Enrico Barausse and Luca Santoni for discussions.

\appendix

\section{Reflection symmetry}
\label{app:reflection_symmetry}

The assumption~\eqref{eq:mirrorModeAmplitude}, commonplace in ringdown models(e.g.~\cite{Berti:2007fi, Berti:2007zu,Yi:2024elj}), is related to reflection symmetry (see also app. B of~\cite{Isi:2021iql}).
Let us show here that eq.~\eqref{eq:mirrorModeAmplitude} comes from assuming that the waveform is reflection symmetric with respect to the plane of the binary, i.e. symmetric under $\theta \rightarrow \pi-\theta$. In this Appendix we will perform all computations in the TT gauge, so we will suppress the TT labels on variables to avoid clutter.
Denoting by $\mathcal{R}^i_j = \delta^i_j - 2 \delta^i_z \delta_j^z$ the matrix of the linear transformation defining reflection symmetry in Cartesian coordinates, it is easy to show that the polarization tensors defined in~\eqref{eq:polarizationTensors} transform as
\begin{align}
    e^+_{ij}(\pi - \theta, \phi) &= R_i^k R_j^l e^+_{kl}(\theta, \phi) \; , \\
    e^\times_{ij}(\pi - \theta, \phi) &= - R_i^k R_j^l e^\times_{kl}(\theta, \phi) \; ,
\end{align}
so that imposing that the waveform is refection symmetric i.e. $h_{ij}(\pi - \theta, \phi) = R_i^k R_j^l h_{kl} (\theta, \phi)$ imposes the transformation properties
\begin{align}
    \mathfrak h_+(\pi-\theta, \phi) &= \mathfrak h_+(\theta, \phi) \label{eq:transfo_hplus} \; , \\
    \mathfrak h_\times(\pi-\theta, \phi) &= - \mathfrak h_\times(\theta, \phi) \label{eq:transfo_hcross} \; .
\end{align}
The second step of the proof is to derive what is the transformation property of a single $(\ell m)$ component of the waveform model~\eqref{eq:NPscalar}. Let us denote $\mathfrak h_+-i\mathfrak h_\times = \sum_{\ell m} \psi_{\ell m} \; \vphantom{a}_{-2}Y_{\ell m}(\theta, \phi)$. Using the property $\vphantom{a}_{-2}Y_{\ell m}(\pi-\theta, \phi) = (-1)^\ell \vphantom{a}_{-2} Y_{\ell -m}^*(\theta, \phi)$ we get that reflection symmetry imposes
\begin{equation}
    \psi_{\ell (-m)}= (-1)^\ell \psi_{\ell m}^* \; .
\end{equation}
This is exactly the transformation needed to find the relation~\eqref{eq:mirrorModeAmplitude} for the mirror mode amplitudes.

\section{A toy model}
\label{app:toy_model}
We can understand the main features of quadratic modes with a simple toy-model. Let us consider the following differential equation on the variable $h(t,r)$:
\begin{equation}
    \frac{\mathrm{d}^2 h}{\mathrm{d}r^2} - \frac{\mathrm{d}^2 h}{\mathrm{d}t^2} = \epsilon h^2 \; ,
\end{equation}
where $\epsilon$ is the small expansion parameter. We can solve it perturbatively by expanding $h=h^{(1)} + \epsilon h^{(2)} + \dots$ At leading order (i.e. setting $\epsilon=0$), the solution for $h^{(1)}$ is a superposition of ingoing and outgoing waves. We will just assume that the spatial boundary conditions of the problem are such that the frequencies $\omega_j$ of these waves are quantized by some integer $j$ so that
\begin{equation}
    h^{(1)} = \sum_j A_j e^{i \omega_j (r-t)} \; .
\end{equation}
Let us now solve for the second-order solution $h^{(2)}$, which obeys the equation
\begin{equation}
      \frac{\mathrm{d}^2 h^{(2)}}{\mathrm{d}r^2} - \frac{\mathrm{d}^2 h^{(2)}}{\mathrm{d}t^2} = \sum_{j,k} A_j A_k e^{i (\omega_j + \omega_k) (r-t)} \; .
\end{equation}
The homogeneous solution of this equation is just a renormalization of the first-order solution $h^{(1)}$. On the other hand, the source term on the right-hand side oscillates at a frequency $\omega_j + \omega_k$ which we assume is not contained in the spectrum of the linear solution, as is the case for QNM (i.e., $\omega_l \neq \omega_j + \omega_k$ for all $j,k,l$). Let us focus on a single frequency component of the source term. Notice that, for a given frequency $\omega_j + \omega_k$, there are in fact two terms in the sum contributing to the source if $j \neq k$, and one if $j=k$. We can easily take this into account by multiplying the source by $2$ and dividing it by a symmetry factor $\texttt S$, where $\texttt S = 2$ if $j=k$ and $\texttt S=1$ otherwise. 

Then, the particular solution for $h^{(2)} = \tilde h^{(2)} e^{-i (\omega_j + \omega_k) t}$ oscillates at the new frequency $\omega_j + \omega_k$ and is given by
\begin{equation}
    \tilde h^{(2)} = B_{jk} e^{i (\omega_j + \omega_k) r} + C_{jk} e^{-i (\omega_j + \omega_k) r} - \frac{i A_j A_k}{\texttt S(\omega_j + \omega_k)} r e^{i (\omega_j + \omega_k) r} \; .
\end{equation}
Several important features emerge from this equation. First, notice that the last term seem to diverge at large radius $r$. This unphysical feature is avoided in the complete calculation that we perform in the main body of the paper by defining a ``regularized'' source term, whose asymptotic behavior will ensure that the QNM amplitude goes to a constant at the spatial boundaries. The computation of the physical gravitational wave amplitude at second order will only involve the regularized source term.

Second, notice that the only freedom that we have at our disposal in order to ensure that $h^{(2)}$ respects the spatial boundary conditions is to tune the free amplitudes $B_{jk}$ and $C_{jk}$, since the frequency $\omega_j + \omega_k$ is not a free parameter like at first order but is instead imposed by the source term. This point highlight the essential difference between first and second-order modes: while at first order the amplitudes of the modes are arbitrary parameters that should be obtained from the initial conditions of the merger (e.g. the masses and spins of the progenitors BH), at second order \textit{both frequencies and amplitudes} are fixed by the spatial boundary conditions, which in the full problem will be the requirement that the wave is purely infalling at the horizon and purely outgoing at infinity. 

\section{Spherical harmonics and angular integrals}
\label{app:tensor_harmonics}
In this appendix we review, following \cite{Spiers:2023mor}, the conventions regarding the decomposition of a spherically symmetric spacetime in time-radial and angular components. We focus on the angular components, in particular on the construction of some tools that will allow us to easily make use of the properties of spherical symmetry. To begin, the Schwarzschild metric can be decomposed as: 
\begin{equation}
    ds^2 = g_{ab} dx^a dx^b + r^2 \Omega_{AB} d\theta^A d\theta^B \; ,
\end{equation}
where, using standard polar coordinates $\theta^A = (\theta, \phi)$ 
\begin{equation}
    \Omega_{AB} := \begin{pmatrix}
        1 & 0 \\
        0 & \sin{\theta}^2
    \end{pmatrix} \; ,
\end{equation}
while the time and radial components in coordinates $(t, r)$ are: 
\begin{equation}
    g_{ab} := \begin{pmatrix}
        -f(r) & 0 \\
        0    & f(r)^{-1}
    \end{pmatrix} \; .
\end{equation}
Focusing on the angular part, it is possible to choose a different basis from the coordinate one, in particular one may work with a set of complex null vectors which forms a complete basis for the real vector space: 
\begin{equation}
    m_A = \left(1, i\sin{\theta} \right) \quad \text{and} \quad m^{*}_A  = \left(1, -i\sin\theta \right) \; ,
\end{equation}
where the star denotes complex conjugation. This new basis will be useful for the definition of the so called \textit{spin-weighted spherical harmonics} which will be used to perform angular integrals. \\

Splitting the $(t, r)$ sector and the angular one, we may define a covariant derivative on the unit sphere. In accordance with the existing literature, we will denote this as $D$. The non-zero Christoffel symbols are: 
\begin{equation}
    \Gamma^{\phi}_{\theta \phi} = \cot{\theta} \; , \qquad \Gamma^{\theta}_{\phi \phi} = -\sin{\theta}\cos{\theta} \; ,
\end{equation}
modulo permutations of lower indices.

\subsection{Tensor harmonics}
Due to the invariance under rotations of the unit sphere $\mathbb{S}^2$, scalar functions defined on the latter can be decomposed in spherical harmonics $Y_{\ell m}(\theta, \phi)$ which are the eigenfunctions of the Laplace operator on $\mathbb{S}^2$: 
\begin{equation}
    D_AD^A Y^{\ell m}(\theta, \phi) = -\ell(\ell+1) Y^{\ell m}(\theta, \phi) \; .
\end{equation}
The same is valid for generic tensorial quantities, with the only difference that the basis for the decomposition will be related to covariant derivatives of spherical harmonics. We begin discussing the decomposition of vectors. We define: 
\begin{equation}
    \begin{split}
        Y_A^{\ell m}(\theta, \phi) & := \partial_A Y^{\ell m} \; , \\
        X_A^{\ell m}(\theta, \phi) & := - \epsilon_A^{\ C} \partial_C Y^{\ell m} \; ,
    \end{split}
\end{equation}
where $\mathbf{\epsilon}$ is the Levi-Civita tensor on the unit sphere, in particular: 
\begin{equation}
    \epsilon_{\theta \phi} = -\epsilon_{\phi\theta} = \sin{\theta} \; .
\end{equation}
These two \textit{vector harmonics} form a basis for vector fields on the sphere, in fact for each point on the sphere $\mathbf{X}$ and $\mathbf{Y}$ are orthogonal to each other. A generic vector field can be therefore decomposed as: 
\begin{equation}
    \mathbf{V}(\theta, \phi) = \sum_{\ell m}  V_+^{\ell m} \mathbf{Y}^{\ell m}(\theta, \phi) + V_-^{\ell m} \mathbf{X}^{\ell m}(\theta, \phi) \; .
\end{equation}
$V_+$ and $V_-$ are the components of $\mathbf{V}$ with respect to the basis defined by $\mathbf{Y}$ and $\mathbf{X}$. The plus and the minus are related to parity properties of vector harmonics. Scalar spherical harmonics have parity $+$ or $-$ depending on $\ell$, in particular
\begin{equation}
    Y^{\ell m}(\pi - \theta, \phi + \pi) = (-1)^\ell Y^{\ell m}(\theta, \phi) \; .
\end{equation}
The same is true for the vector harmonics: 
\begin{equation}
    \begin{split}
        \mathbf{Y}^{\ell m}(\theta, \phi) &  \rightarrow (-1)^{\ell}\mathbf{Y}^{\ell m}(\theta, \phi) \; , \\
        \mathbf{X}^{\ell m}(\theta, \phi) &  \rightarrow (-1)^{\ell+1}\mathbf{X}^{\ell m}(\theta, \phi) \; ,
    \end{split}
\end{equation}
which can be easily seen writing the vectors in the $(\theta, \phi)$ coordinate base and performing the transformation $(\theta, \phi)\rightarrow(\pi-\theta, \phi+ \pi)$ accounting also for the change of basis. In other words, $\mathbf{Y}$ and $\mathbf{X}$ are respectively a set of vectors and pseudo-vectors indexed by $\ell$ and $m$ and this justifies the notation $+$ and $-$ for the decomposition of $\mathbf{V}$. \\

The same considerations can be extended to higher rank tensors. The irreducible representations of a generic two indices tensor under $SO(3)$ are a symmetric, anti-symmetric and trace part. Using the previous machinery, it is immediate to get a basis for two indices \textit{symmetric} tensors that makes parity explicit: 
\begin{equation}
    \begin{split}
            Y^{\ell m}_{AB} & := D_{\left < A \right.}D_{\left. B \right>} Y^{\ell m} \equiv D_AD_BY^{\ell m} - \frac{1}{2}D_AD^A Y^{\ell m} \; ,\\
            X^{\ell m}_{AB} & := - \epsilon_{(A}^{\ C} D_{B)} D_C Y^{\ell m} \; ,
    \end{split}
\end{equation}
where $\left< \cdot \right>$ denotes the symmetrized traceless component. These tensors are called \textit{tensor spherical harmonics}, and as in the previous case have respectively parity $+$ and $-$ on top of the parity given by $\ell$. A rank two \textit{symmetric} tensor can therefore be decomposed as: 
\begin{equation}
    t_{AB} = \sum_{\ell m} t_{\circ}^{\ell m} \Omega_{AB} Y^{\ell m} + t_{+}^{\ell m} Y_{AB}^{\ell m} + t_{-}^{\ell m} X_{AB}^{\ell m}  \; .
\end{equation}
Again, the plus and minus make explicit the parity of the associated tensor harmonic. Notice that decomposing a \textit{generic} tensor would also require the definition of two anti-symmetric tensors with definite parity. This explains the decomposition in equations (\ref{eq:spherical_decomposition1},~\ref{eq:spherical_decomposition2},~\ref{eq:spherical_decomposition3}) if the time-radial components are thought as indexing scalars of $SO(3)$. \\

To make contact with common conventions in the literature it may be useful to write explicitly the matrices $Y^{\ell m}_{AB}$ and $X^{\ell m}_{AB}$: 
\begin{equation}
    Y^{\ell m}_{AB} = \begin{pmatrix}
        \partial^2_\theta Y^{\ell m} +\frac{1}{2}\ell(\ell+1)Y^{\ell m} & -\cot\theta \partial_\phi Y^{\ell m} + \partial_\theta\partial_\phi Y^{\ell m} \\
            -\cot\theta \partial_\phi Y^{\ell m} + \partial_\theta\partial_\phi Y^{\ell m} & \partial^2_\phi Y^{\ell m} +\cos\theta\sin\theta\partial_\theta Y^{\ell m} +\frac{1}{2} \ell(\ell+1)\sin^2\theta Y^{\ell m}
    \end{pmatrix} \; ,
\end{equation}
\begin{equation}
    X^{\ell m}_{AB} = \begin{pmatrix}
        \frac{1}{\sin\theta}\left(\cot\theta \partial_\phi Y -\partial_\theta\partial_\phi Y\right) & \sin\theta (\partial_\theta^2Y+\frac{1}{2}\ell(\ell+1)Y) \\
            \sin\theta (\partial_\theta^2Y+\frac{1}{2}\ell(\ell+1)Y) & -\sin\theta\left(\cot\theta \partial_\phi Y -\partial_\theta\partial_\phi Y\right)
    \end{pmatrix} \; .
\end{equation}
All the spherical tensors we defined are orthogonal under the scalar product defined on the sphere. Our definitions can be compared for example with eq.~(3.2-3.11) in \cite{Nakano:2007cj}, where they appear with an extra prefactor that ensures orthonormality.

\subsection{Spin-weighted spherical harmonics}
In the paper we made use of the machinery of spin-weighted spherical harmonics in order to compute the integrals that are necessary to extract the spherical components of energy-momentum tensor generated by the first order perturbation. A spin-weighted spherical harmonic is a function of the usual spherical harmonics defined as: 
\begin{equation}
\label{eq:spin-weighted_harmonics}
    \vphantom{|}_{s}Y^{\ell m} = \frac{1}{\lambda_s}\begin{cases}
        (-1)^s \eth^s Y^{\ell m} & 0 \le s \le \ell  \\ 
        \eth^{*|s|} Y^{\ell m} &  -\ell \le s \le 0
    \end{cases} \; ,
\end{equation}
where $\lambda_s$ was defined in \cref{eq:def:lambda_mu_Lambda} and the differential operators act as: 
\begin{align}
\label{eq:spin-weighted_differential_op}
    \begin{cases}
        \eth \  \vphantom{|}_{s} Y_{\ell m} & = \left(m^AD_A -s D_Am^A \right)\  \vphantom{|}_{s} Y_{\ell m}   \\
        \eth^* \  \vphantom{|}_{s} Y_{\ell m} & = \left(m^{A*}D_A +s D_Am^{A*} \right)\ \vphantom{|}_{s} Y_{\ell m}
    \end{cases} \; .
\end{align}
The vector and tensor spherical harmonics can then be expressed in terms of spin-weighted spherical harmonics as
\begin{align}
    Y_A^{\ell m} &= \frac{\lambda_1}{2} \big({}_{-1}Y^{\ell m} m_A -  {}_{1}Y^{\ell m} m_A^*\big) \label{eq:YA_from_1Y}  \; , \\
    X_A^{\ell m} &= -i \frac{\lambda_1}{2} \big({}_{-1}Y^{\ell m} m_A +  {}_{1}Y^{\ell m} m_A^*\big) \label{eq:XA_from_1Y} \; , \\
    Y_{AB}^{\ell m} &= \frac{\lambda_2}{4} \big({}_{-2}Y^{\ell m} m_A m_B +  {}_{2}Y^{\ell m} m_A^* m_B^*\big) \label{eq:YAB_from_2Y} \; , \\
    X_{AB}^{\ell m} &= - i \frac{\lambda_2}{4} \big({}_{-2}Y^{\ell m} m_A m_B -  {}_{2}Y^{\ell m} m_A^* m_B^*\big) \label{eq:XAB_from_2Y} \; .
\end{align}
and similarly for higher indices tensors.
\subsection{Angular integrals}
\label{app:angular_integrals}
When computing the source term $S_{\mu\nu}$ defined in \cref{eq:EinstEqSecondOrder} (projected on the basis of spherical tensors) one has to compute angular integrals of products of three spherical tensors.

To perform such integrals, the (well known) strategy that we employed was to write any component of any spherical tensor using derivatives of scalar spherical harmonics according to \cref{eq:YA_from_1Y,eq:XA_from_1Y,eq:YAB_from_2Y,eq:XAB_from_2Y} up to trigonometric functions. Replacing such derivative terms with spin-weighted harmonics, one obtains expressions involving the product of three spin-weighted harmonics (without derivatives) and many trigonometric functions. Because ultimately we are always dealing with scalars (the components of $S_{\mu\nu}$ stripped of their angular dependence), all the trigonometric functions simplify and the integrals always reduce to the form
\begin{equation}
\label{eq:generalized_CB_coefficients_definition}
    C_{\ell_1 m_1s_1,\ell_2 m_2s_2,\ell ms}=\int \vphantom{|}_{s_1}Y^{\ell_1 m_1}(\theta, \phi)\,\vphantom{|}_{s_2}Y^{\ell_2 m_2}(\theta, \phi)\,\vphantom{|}_{s}Y^{\ell m}(\theta, \phi) \;\sin\theta\dd\theta\dd\phi \; ,
\end{equation}
This integral can be expressed using Wigner 3-$j$ symbols (closely related to Clebsh-Gordan coefficients) and gives
\begin{equation}
\label{eq:generalized_CB_coefficients_value}
    C_{\ell_1 m_1s_1,\ell_2 m_2s_2,\ell ms}=\sqrt{\frac{(2\ell_1+1)(2\ell_2+1)(2\ell+1)}{4\pi}}
    \begin{pmatrix}
        \ell_1&\ell_2&\ell\\
        -s_1&-s_2&-s
    \end{pmatrix}
    \begin{pmatrix}
        \ell_1&\ell_2&\ell\\
        m_1&m_2&m
    \end{pmatrix} \; .
\end{equation}
This result allows us to perform all the integrals we face. To be more precise, given we want to project $S_{\mu\nu}$ onto a basis of tensor harmonics, our integrals will involve the complex conjugates $Y^{\ell,m\,*},\,Y_A^{\ell,m\,*},\,etc$. We get rid of complex conjugation by flipping the sign of $m$ and $s$ and multiplying by $(-1)^{s+m}$, and then we proceed as described above.

\section{Notation}
\label{app:notation_literature}
While we followed the notation of Spiers, Pound and Wardell~\cite{Spiers:2023mor}, we also found it helpful to cross-checked some of our intermediate results with other authors. Hence, in this appendix we compare our notation with different choices that appeared in the literature.

The regularity problems of the master scalars for QNM solutions were recognized and solved in a particular case by Ioka and Nakano in~\cite{Nakano:2007cj}.
Their conventions for the metric and the source term are summarized in \cref{tab:notation_comparisons}.

\begin{table}[h]
    \centering
    \begin{tabular}{|c|c|c|c|}
        \hline
        $h_{\mu\nu}$ (Us) & $h_{\mu\nu}$ (Ioka, Nakano) & $S_{\mu\nu}$ (Us) & $S_{\mu\nu}$ (Ioka, Nakano) \\
        \hline
         $h_{tt}$ & $f(r)H_0$ & $S_{tt}$ & $8\pi \mathcal{A}_0$ \\
         $h_{tr}$ & $H_1$ & $S_{tr}$ & $8\pi\frac{i}{\sqrt{2}} \mathcal{A}_1$ \\
         $h_{rr}$ & $\frac{1}{f(r)} H_2$ & $S_{rr}$ & $8\pi \mathcal{A}$ \\
         $h_{\circ}$ & $r^2 K -\frac{\ell(\ell+1)}{2}r^2 G$ & $S_{\circ}$ &  $\frac{8\pi r^2}{\sqrt{2}} \mathcal{G}$\\
         $h_{t+}$ &$h_0^{(e)}$ & $S_{t+}$ &  $\frac{8\pi i r}{\sqrt{2\ell(\ell+1)}} \mathcal{B}_0$\\
         $h_{r+}$ & $ h_1^{(e)}$ &  $S_{r+}$ & $\frac{8\pi r}{\sqrt{2\ell(\ell+1)}} \mathcal{B}$  \\
         $h_{t-}$ & $ h_0 $& $S_{t-}$ &$-\frac{8\pi r}{\sqrt{2\ell(\ell+1)}} \mathcal{Q}_0$   \\
         $h_{r-}$ & $h_1 $ & $S_{r-}$ &$-\frac{8\pi i r}{\sqrt{2\ell(\ell+1)}} \mathcal{Q}$   \\
         $h_{+}$ & $r^2G$ & $S_{+}$ & $\frac{ 16\pi r^2}{\sqrt{2\ell(\ell+1)(\ell-1)(\ell+2)}} \mathcal{F}$ \\
         $h_{-}$ & $-i h_2$ & $S_{-}$ & $ -\frac{16\pi i r^2}{\sqrt{2\ell(\ell+1)(\ell-1)(\ell+2)}} \mathcal{D}$  \\
         \hline
    \end{tabular}
    \caption{Comparison of notations for the decomposition of $h_{\mu\nu}$ and $S_{\mu\nu}$ between us and Ioka and Nakano.}
    \label{tab:notation_comparisons}
\end{table}
Ioka and Nakano use our same Zerilli master scalar at linear order, but at second order their $\chi$ is related to our $\psi^{(2)}$ by
\begin{align}
    \chi = \frac{r f(r)}{\lambda_1 r+3M} \left[ \frac{-i\omega  h_\circ}{r f(r)} - h_{tr} \right] = \\
    = \frac{r f(r)}{\lambda_1 r +3M} \left[ -i\omega r\partial_r\psi^{(2)} -i\frac{\omega A}{rf(r)} \psi^{(2)} +i\omega\frac{4r^4}{\lambda_1^2 \Lambda r f(r)} S_{tt} +i\omega r\partial_r\psi^{(2)} +\right.\\
    \left.+i\omega rB \psi^{(2)} -\frac{2r^2}{\lambda_1^2}\left( S_{tr} +i\frac{2\omega r}{\Lambda f(r)}  S_{tt} \right) \right] = \\
    = \frac{r f(r)}{\lambda_1 r +3M} \left[ -i\omega\left( \frac{A}{r f(r)} - r B \right)\psi^{(2)} -\frac{2r^2}{\lambda_1^2} S_{tr} \right] =
    -i\omega \psi^{(2)} -\frac{4 r^2 f(r)}{\Lambda \lambda_1^2} S_{tr}
\end{align}

Finally, Brizuela in~\cite{Brizuela:2009qd} works with a re-scaled version of our master scalar:
\begin{align}
    \mathbf \Psi = -\frac{2f}{\Lambda}\left(-rfh_{rr}+r^2(r^{-2}h_\circ)'\right)+\frac{1}{r}h_\circ
    = \frac{\lambda_{1}^2}{2}\psi_+\\
    \Pi = -\epsilon^{ab} \frac{1}{r^2} \delta_b h_{a-} + \frac{2}{r^3} \epsilon^{ab} r_b h_{a-} = - \frac{2r^3}{\mu_\ell^2} \psi_-
\end{align}
where the comparison is performed using on both sides the RW gauge.

\section{Code computing the source term}
\label{app:code}

In this appendix, we give a brief account of the \textsc{Mathematica} codes \textquotedblleft\emph{Asymptotics}\textquotedblright{} we used to compute the source term $\mathfrak S$ defined in \cref{eq:regulated_RWZ_equation}. We separate the computation according to the parity of the modes involved (Regge-Wheeler or Zerilli sectors), both at linear and at quadratic order, for a total of six possible combinations, as displayed in \cref{eq:parity_sum_Zerilli,eq:parity_sum_RW}. These codes also report the metric reconstruction at large $r$ in RW gauge starting from the regulated master scalar, which displays many nontrivial cancellations of divergent terms.

We also wrote a code that, starting from the Einstein equations, derives \cref{eq:RWZ_equation,eq:master_scalar_even,eq:master_scalar_odd,eq:metric_reconstruction}. This code is quite straightforward, so we will not comment further, except to point out that \cref{eq:RWZ_equation} can be derived from the Einstein equations, as is well known, without relying on an explicit expression for $S_{\mu\nu}$, but rather only on its conservation. While reading our notebooks, the reader should note that, for convenience, our code defines $\mu$ and $\lambda_s$ in \cref{eq:def:lambda_mu_Lambda} without the square root.

\subsection{Computing $S_{\mu\nu}$}
We begin by computing the $S_{\mu\nu}$ tensor appearing in \cref{eq:EinstEqSecondOrder}. We do this by computing, using the Black Hole Perturbation Toolkit, the Einstein tensor of the metric
\begin{equation}
    g=\bar g + \epsilon\left( \sigma_1 e\, h_{1,+}^{(1)} + \sigma_1 o\, h_{1,-}^{(1)} + \sigma_2 e\, h_{2,+}^{(1)} + \sigma_2 o\, h_{1,-}^{(1)} \right)+\OO(\epsilon^3)
\end{equation}
which gives us $G^{(2)}[h^{(1)},h^{(1)}]$. In the equation above, all $h$ terms have the appropriate time and angular dependence which we don't write explicitly here. The dummy parameters $\sigma_{1,2}$ are introduced to identify the contributions that couple the two perturbations (in contrast to self-coupling), which are proportional to $\sigma_1\times\sigma_2$. Similarly, the parameters $(e,\,o)$ are introduced to help single out the desired parity of the linear modes. The $\OO(\epsilon^2)$ term was omitted because we had already verified that it gives $G^{(1)}[h^{(2)}]$ in \cref{eq:EinstEqSecondOrder}.
The linear metric perturbations are then expressed in terms of the master scalars in \cref{eq:master_scalar_even,eq:master_scalar_odd}.

\subsection{Angular integrals}
While we already discussed the theory behind angular integrals of spin-weighted spherical harmonics in \cref{app:angular_integrals}, here we give some more details on the code implementation.

Our code converts derivatives of spherical harmonics to spin-weighted spherical harmonics using \cref{eq:YA_from_1Y,eq:XA_from_1Y,eq:YAB_from_2Y,eq:XAB_from_2Y}, and then performs the integrals using \cref{eq:generalized_CB_coefficients_definition}. We exploit the symmetry of the 3j symbols
\begin{equation}
    \begin{pmatrix}
        \ell_1&\ell_2&\ell\\
        s_1&s_2&s
    \end{pmatrix} = (-1)^{\ell_1+\ell_2+\ell} \begin{pmatrix}
        \ell_1&\ell_2&\ell\\
        -s_1&-s_2&-s
    \end{pmatrix}
\end{equation}
to make $s_1\ge 0$ and, in case $s_1=0$, $s_2\ge0$. This allows us to collect terms and work with simpler expressions. We note that $(-1)^{\ell_1+\ell_2+\ell}$ is completely fixed by the parity of the modes, so we hardcode it keeping in mind that the source obtained in this way will be correct \emph{only} for processes not forbidden by parity (see \cref{eq:selection_rule_parity}).

\subsection{Regulation of $S$}
The source term $s(r)$ defined in \cref{eq:sourceSs} gives a regular master scalar if it vanishes at least linearly at the horizon and at least as $1/r^2$ at infinity. To perform the required asymptotic expansions, we computed the ratio $\psi'(r)/\psi(r)$ to high order in $(r-2M)$ and in $1/r$ for both the even and the odd sectors, using asymptotic expansions in the RW and Zerilli equations. We are then able to compute $s(r)$ close to the horizon, finding that it is always regular regardless of the parity sector.

On the contrary, we find a divergent behaviour at large $r$, in particular the function $\Delta(r)$ defined in \cref{eq:regulated_master_scalar} is chosen depending on the parities according to the ansatze in \cref{tab:Delta_ansatz},
\begin{table}[h]
\centering
\begin{tabular}{|c|c|c|c|c|}
 \hline
 Linear mode 1    & Linear mode 2    & Quadratic mode    & $\Delta(r)$    \\
 \hline
 \multirow{2}*{Even}
            & \multirow{2}*{Even}
                        & Even    & $a_2 r^2 +a_1 r$ \\
                        
                        && Odd & $/$ \\
 \hline
 \multirow{2}*{Even}
            & \multirow{2}*{Odd}
                        & Even    & $a_1 r$ \\
                        
                        && Odd & $a_1 r$ \\
 \hline
 \multirow{2}*{Odd}
            & \multirow{2}*{Odd}
                        & Even    & $a_1 r$ \\
                        
                        && Odd & $/$ \\
\hline
\end{tabular}
\caption{The ansatz chosen for $\Delta(r)$, which depends on the parity of the linear and quadratic modes.}
\label{tab:Delta_ansatz}
\end{table}
where $/$ indicates that $s(r)$ is already regular, and the $c_i$ vary depending on the parities. It is worth mentioning that, using these ansatze, the source term $\mathfrak s(r)$ receives corrections up to order $\OO(r)$ from $a_2$ and $\OO(1)$ from $a_1$; perhaps surprisingly, the $\OO(1/r)$ term is always automatically cancelled. The reason for this simplification is that the $\OO(1/r^2)$ behaviour expected from a regular source comes from imposing that $\Psi(r)$ in \cref{eq:regulated_master_scalar} should go to a pure exponential at large $r$ (regular behaviour). Given the regularity of the linear master scalars $\psi^{(1)}_{1,2}$, and assuming a power law expansion in $r$ and $1/r$ for $\psi^{(2)}$, our ansatze for $\Delta(r)$ can already eliminate all divergences from $\Psi$, which requires $\mathfrak s(r)$ to be quadratically vanishing. An alternative point of view is that to cancel a possible $\OO(1/r)$ term in $\mathfrak s(r)$, our $\Delta(r)$ would have to contain a term $\propto \log(r)/r$, which would later spoil the finiteness of the observables at large $r$.

\bibliographystyle{utphys}
\bibliography{bib}

\providecommand{\href}[2]{#2}\begingroup\raggedright\begin{thebibliography}{10}

\bibitem{Regge:1957td}
T.~Regge and J.~A. Wheeler, ``{Stability of a Schwarzschild singularity},'' \href{http://dx.doi.org/10.1103/PhysRev.108.1063}{{\em Phys.Rev.} {\bfseries 108} (1957) 1063--1069}.

\bibitem{Zerilli:1970aa}
F.~J. Zerilli, ``Gravitational field of a particle falling in a schwarzschild geometry analyzed in tensor harmonics,'' \href{http://dx.doi.org/10.1103/PhysRevD.2.2141}{{\em Physical Review D} {\bfseries 2} no.~10, (1970) 2141--2160}.

\bibitem{1973ApJ...185..635T}
S.~A. {Teukolsky}, ``{Perturbations of a Rotating Black Hole. I. Fundamental Equations for Gravitational, Electromagnetic, and Neutrino-Field Perturbations},'' \href{http://dx.doi.org/10.1086/152444}{{\em \apj} {\bfseries 185} (Oct., 1973) 635--648}.

\bibitem{Blanchet:2013haa}
L.~Blanchet, ``{Gravitational Radiation from Post-Newtonian Sources and Inspiralling Compact Binaries},'' \href{http://dx.doi.org/10.12942/lrr-2014-2}{{\em Living Rev. Rel.} {\bfseries 17} (2014) 2},
\href{http://arxiv.org/abs/1310.1528}{{\ttfamily arXiv:1310.1528 [gr-qc]}}.

\bibitem{Berti:2009kk}
E.~Berti, V.~Cardoso, and A.~O. Starinets, ``{Quasinormal modes of black holes and black branes},'' \href{http://dx.doi.org/10.1088/0264-9381/26/16/163001}{{\em Class. Quant. Grav.} {\bfseries 26} (2009) 163001}, \href{http://arxiv.org/abs/0905.2975}{{\ttfamily arXiv:0905.2975 [gr-qc]}}.

\bibitem{Konoplya:2011qq}
R.~A. Konoplya and A.~Zhidenko, ``{Quasinormal modes of black holes: From astrophysics to string theory},'' \href{http://dx.doi.org/10.1103/RevModPhys.83.793}{{\em Rev. Mod. Phys.} {\bfseries 83} (2011) 793--836}, \href{http://arxiv.org/abs/1102.4014}{{\ttfamily arXiv:1102.4014 [gr-qc]}}.

\bibitem{Ghosh:2021mrv}
A.~Ghosh, R.~Brito, and A.~Buonanno, ``{Constraints on quasinormal-mode frequencies with LIGO-Virgo binary\textendash{}black-hole observations},'' \href{http://dx.doi.org/10.1103/PhysRevD.103.124041}{{\em Phys. Rev. D} {\bfseries 103} no.~12, (2021) 124041}, \href{http://arxiv.org/abs/2104.01906}{{\ttfamily arXiv:2104.01906 [gr-qc]}}.

\bibitem{LIGOScientific:2016lio}
{\bfseries LIGO Scientific, Virgo} Collaboration, B.~P. Abbott {\em et~al.}, ``{Tests of general relativity with GW150914},'' \href{http://dx.doi.org/10.1103/PhysRevLett.116.221101}{{\em Phys. Rev. Lett.} {\bfseries 116} no.~22, (2016) 221101}, \href{http://arxiv.org/abs/1602.03841}{{\ttfamily arXiv:1602.03841 [gr-qc]}}. [Erratum: Phys.Rev.Lett. 121, 129902 (2018)].

\bibitem{LIGOScientific:2020tif}
{\bfseries LIGO Scientific, Virgo} Collaboration, R.~Abbott {\em et~al.}, ``{Tests of general relativity with binary black holes from the second LIGO-Virgo gravitational-wave transient catalog},'' \href{http://dx.doi.org/10.1103/PhysRevD.103.122002}{{\em Phys. Rev. D} {\bfseries 103} no.~12, (2021) 122002}, \href{http://arxiv.org/abs/2010.14529}{{\ttfamily arXiv:2010.14529 [gr-qc]}}.

\bibitem{Buonanno:2006ui}
A.~Buonanno, G.~B. Cook, and F.~Pretorius, ``{Inspiral, merger and ring-down of equal-mass black-hole binaries},'' \href{http://dx.doi.org/10.1103/PhysRevD.75.124018}{{\em Phys. Rev. D} {\bfseries 75} (2007) 124018}, \href{http://arxiv.org/abs/gr-qc/0610122}{{\ttfamily arXiv:gr-qc/0610122}}.

\bibitem{Berti:2007fi}
E.~Berti, V.~Cardoso, J.~A. Gonzalez, U.~Sperhake, M.~Hannam, S.~Husa, and B.~Bruegmann, ``{Inspiral, merger and ringdown of unequal mass black hole binaries: A Multipolar analysis},'' \href{http://dx.doi.org/10.1103/PhysRevD.76.064034}{{\em Phys. Rev. D} {\bfseries 76} (2007) 064034}, \href{http://arxiv.org/abs/gr-qc/0703053}{{\ttfamily arXiv:gr-qc/0703053}}.

\bibitem{Berti:2007zu}
E.~Berti, J.~Cardoso, V.~Cardoso, and M.~Cavaglia, ``{Matched-filtering and parameter estimation of ringdown waveforms},'' \href{http://dx.doi.org/10.1103/PhysRevD.76.104044}{{\em Phys. Rev. D} {\bfseries 76} (2007) 104044}, \href{http://arxiv.org/abs/0707.1202}{{\ttfamily arXiv:0707.1202 [gr-qc]}}.

\bibitem{Baibhav:2017jhs}
V.~Baibhav, E.~Berti, V.~Cardoso, and G.~Khanna, ``{Black Hole Spectroscopy: Systematic Errors and Ringdown Energy Estimates},'' \href{http://dx.doi.org/10.1103/PhysRevD.97.044048}{{\em Phys. Rev. D} {\bfseries 97} no.~4, (2018) 044048}, \href{http://arxiv.org/abs/1710.02156}{{\ttfamily arXiv:1710.02156 [gr-qc]}}.

\bibitem{Giesler:2019uxc}
M.~Giesler, M.~Isi, M.~A. Scheel, and S.~Teukolsky, ``{Black Hole Ringdown: The Importance of Overtones},'' \href{http://dx.doi.org/10.1103/PhysRevX.9.041060}{{\em Phys. Rev. X} {\bfseries 9} no.~4, (2019) 041060}, \href{http://arxiv.org/abs/1903.08284}{{\ttfamily arXiv:1903.08284 [gr-qc]}}.

\bibitem{Cheung:2023vki}
M.~H.-Y. Cheung, E.~Berti, V.~Baibhav, and R.~Cotesta, ``{Extracting linear and nonlinear quasinormal modes from black hole merger simulations},'' \href{http://arxiv.org/abs/2310.04489}{{\ttfamily arXiv:2310.04489 [gr-qc]}}.

\bibitem{Barausse:2020rsu}
E.~Barausse {\em et~al.}, ``{Prospects for Fundamental Physics with LISA},'' \href{http://dx.doi.org/10.1007/s10714-020-02691-1}{{\em Gen. Rel. Grav.} {\bfseries 52} no.~8, (2020) 81}, \href{http://arxiv.org/abs/2001.09793}{{\ttfamily arXiv:2001.09793 [gr-qc]}}.

\bibitem{reitze2019cosmic}
D.~Reitze, R.~X. Adhikari, S.~Ballmer, B.~Barish, L.~Barsotti, G.~Billingsley, D.~A. Brown, Y.~Chen, D.~Coyne, R.~Eisenstein, M.~Evans, P.~Fritschel, E.~D. Hall, A.~Lazzarini, G.~Lovelace, J.~Read, B.~S. Sathyaprakash, D.~Shoemaker, J.~Smith, C.~Torrie, S.~Vitale, R.~Weiss, C.~Wipf, and M.~Zucker, ``Cosmic explorer: The u.s. contribution to gravitational-wave astronomy beyond ligo,'' 2019.

\bibitem{punturo:hal-00629986}
M.~e.~a. Punturo, ``{The Einstein Telescope: a third-generation gravitational wave observatory},'' \href{http://dx.doi.org/10.1088/0264-9381/27/19/194002}{{\em {Classical and Quantum Gravity}} {\bfseries 27} no.~19, (Oct., 2010) 194002}. \url{https://hal.archives-ouvertes.fr/hal-00629986}.

\bibitem{Yi:2024elj}
S.~Yi, A.~Kuntz, E.~Barausse, E.~Berti, M.~H.-Y. Cheung, K.~Kritos, and A.~Maselli, ``{Nonlinear quasinormal mode detectability with next-generation gravitational wave detectors},''  (3, 2024) , \href{http://arxiv.org/abs/2403.09767}{{\ttfamily arXiv:2403.09767 [gr-qc]}}.

\bibitem{Ma_2022}
S.~Ma, K.~Mitman, L.~Sun, N.~Deppe, F.~H{\'{e}}bert, L.~E. Kidder, J.~Moxon, W.~Throwe, N.~L. Vu, and Y.~Chen, ``Quasinormal-mode filters: A new approach to analyze the gravitational-wave ringdown of binary black-hole mergers,'' \href{http://dx.doi.org/10.1103/physrevd.106.084036}{{\em Physical Review D} {\bfseries 106} no.~8, (Oct, 2022) }. \url{https://doi.org/10.1103%2Fphysrevd.106.084036}.

\bibitem{London_2014}
L.~London, D.~Shoemaker, and J.~Healy, ``Modeling ringdown: Beyond the fundamental quasinormal modes,'' \href{http://dx.doi.org/10.1103/physrevd.90.124032}{{\em Physical Review D} {\bfseries 90} no.~12, (Dec, 2014) }. \url{https://doi.org/10.1103%2Fphysrevd.90.124032}.

\bibitem{Mitman:2022qdl}
K.~Mitman {\em et~al.}, ``{Nonlinearities in Black Hole Ringdowns},'' \href{http://dx.doi.org/10.1103/PhysRevLett.130.081402}{{\em Phys. Rev. Lett.} {\bfseries 130} no.~8, (2023) 081402}, \href{http://arxiv.org/abs/2208.07380}{{\ttfamily arXiv:2208.07380 [gr-qc]}}.

\bibitem{Cheung:2022rbm}
M.~H.-Y. Cheung {\em et~al.}, ``{Nonlinear Effects in Black Hole Ringdown},'' \href{http://dx.doi.org/10.1103/PhysRevLett.130.081401}{{\em Phys. Rev. Lett.} {\bfseries 130} no.~8, (2023) 081401}, \href{http://arxiv.org/abs/2208.07374}{{\ttfamily arXiv:2208.07374 [gr-qc]}}.

\bibitem{Khera:2023oyf}
N.~Khera, A.~Ribes~Metidieri, B.~Bonga, X.~Jim\'enez~Forteza, B.~Krishnan, E.~Poisson, D.~Pook-Kolb, E.~Schnetter, and H.~Yang, ``{Nonlinear Ringdown at the Black Hole Horizon},'' \href{http://dx.doi.org/10.1103/PhysRevLett.131.231401}{{\em Phys. Rev. Lett.} {\bfseries 131} no.~23, (2023) 231401}, \href{http://arxiv.org/abs/2306.11142}{{\ttfamily arXiv:2306.11142 [gr-qc]}}.

\bibitem{Zhu:2024rej}
H.~Zhu {\em et~al.}, ``{Nonlinear Effects In Black Hole Ringdown From Scattering Experiments I: spin and initial data dependence of quadratic mode coupling},'' \href{http://arxiv.org/abs/2401.00805}{{\ttfamily arXiv:2401.00805 [gr-qc]}}.

\bibitem{Redondo-Yuste:2023seq}
J.~Redondo-Yuste, G.~Carullo, J.~L. Ripley, E.~Berti, and V.~Cardoso, ``{Spin dependence of black hole ringdown nonlinearities},'' \href{http://arxiv.org/abs/2308.14796}{{\ttfamily arXiv:2308.14796 [gr-qc]}}.

\bibitem{Gleiser:1995gx}
R.~J. Gleiser, C.~O. Nicasio, R.~H. Price, and J.~Pullin, ``{Second order perturbations of a Schwarzschild black hole},'' \href{http://dx.doi.org/10.1088/0264-9381/13/10/001}{{\em Class. Quant. Grav.} {\bfseries 13} (1996) L117--L124}, \href{http://arxiv.org/abs/gr-qc/9510049}{{\ttfamily arXiv:gr-qc/9510049}}.

\bibitem{Nicasio_1999}
C.~O. Nicasio, R.~J. Gleiser, R.~H. Price, and J.~Pullin, ``Collision of boosted black holes: Second order close limit calculations,'' \href{http://dx.doi.org/10.1103/physrevd.59.044024}{{\em Physical Review D} {\bfseries 59} no.~4, (Jan, 1999) }. \url{https://doi.org/10.1103%2Fphysrevd.59.044024}.

\bibitem{Gleiser_2000}
R.~J. Gleiser, C.~O. Nicasio, R.~H. Price, and J.~Pullin, ``Gravitational radiation from schwarzschild black holes: the second-order perturbation formalism,'' \href{http://dx.doi.org/10.1016/s0370-1573(99)00048-4}{{\em Physics Reports} {\bfseries 325} no.~2, (Feb, 2000) 41--81}. \url{https://doi.org/10.1016%2Fs0370-1573%2899%2900048-4}.

\bibitem{Brizuela:2006ne}
D.~Brizuela, J.~M. Martin-Garcia, and G.~A. Mena~Marugan, ``{Second and higher-order perturbations of a spherical spacetime},'' \href{http://dx.doi.org/10.1103/PhysRevD.74.044039}{{\em Phys. Rev. D} {\bfseries 74} (2006) 044039}, \href{http://arxiv.org/abs/gr-qc/0607025}{{\ttfamily arXiv:gr-qc/0607025}}.

\bibitem{Brizuela:2007zza}
D.~Brizuela, J.~M. Martin-Garcia, and G.~A.~M. Marugan, ``{High-order gauge-invariant perturbations of a spherical spacetime},'' \href{http://dx.doi.org/10.1103/PhysRevD.76.024004}{{\em Phys. Rev. D} {\bfseries 76} (2007) 024004}, \href{http://arxiv.org/abs/gr-qc/0703069}{{\ttfamily arXiv:gr-qc/0703069}}.

\bibitem{Brizuela:2009qd}
D.~Brizuela, J.~M. Martin-Garcia, and M.~Tiglio, ``{A Complete gauge-invariant formalism for arbitrary second-order perturbations of a Schwarzschild black hole},'' \href{http://dx.doi.org/10.1103/PhysRevD.80.024021}{{\em Phys. Rev. D} {\bfseries 80} (2009) 024021}, \href{http://arxiv.org/abs/0903.1134}{{\ttfamily arXiv:0903.1134 [gr-qc]}}.

\bibitem{Nakano:2007cj}
H.~Nakano and K.~Ioka, ``{Second Order Quasi-Normal Mode of the Schwarzschild Black Hole},'' \href{http://dx.doi.org/10.1103/PhysRevD.76.084007}{{\em Phys. Rev. D} {\bfseries 76} (2007) 084007}, \href{http://arxiv.org/abs/0708.0450}{{\ttfamily arXiv:0708.0450 [gr-qc]}}.

\bibitem{Ioka:2007ak}
K.~Ioka and H.~Nakano, ``{Second and higher-order quasi-normal modes in binary black hole mergers},'' \href{http://dx.doi.org/10.1103/PhysRevD.76.061503}{{\em Phys. Rev. D} {\bfseries 76} (2007) 061503}, \href{http://arxiv.org/abs/0704.3467}{{\ttfamily arXiv:0704.3467 [astro-ph]}}.

\bibitem{Ma:2024qcv}
S.~Ma and H.~Yang, ``{The excitation of quadratic quasinormal modes for Kerr black holes},''  (1, 2024) , \href{http://arxiv.org/abs/2401.15516}{{\ttfamily arXiv:2401.15516 [gr-qc]}}.

\bibitem{Spiers:2023mor}
A.~Spiers, A.~Pound, and B.~Wardell, ``{Second-order perturbations of the Schwarzschild spacetime: practical, covariant and gauge-invariant formalisms},'' \href{http://arxiv.org/abs/2306.17847}{{\ttfamily arXiv:2306.17847 [gr-qc]}}.

\bibitem{Wardell:2021fyy}
B.~Wardell, A.~Pound, N.~Warburton, J.~Miller, L.~Durkan, and A.~Le~Tiec, ``{Gravitational waveforms for compact binaries from second-order self-force theory},'' \href{http://arxiv.org/abs/2112.12265}{{\ttfamily arXiv:2112.12265 [gr-qc]}}.

\bibitem{Bourg:2024jme}
P.~Bourg, R.~Panosso~Macedo, A.~Spiers, B.~Leather, B.~Bonga, and A.~Pound, ``{Quadratic quasi-normal mode dependence on linear mode parity},'' \href{http://arxiv.org/abs/2405.10270}{{\ttfamily arXiv:2405.10270 [gr-qc]}}.

\bibitem{BenAchour:2024skv}
J.~Ben~Achour and H.~Roussille, ``{Quadratic perturbations of the Schwarzschild black hole: The algebraically special sector},'' \href{http://arxiv.org/abs/2406.08159}{{\ttfamily arXiv:2406.08159 [gr-qc]}}.

\bibitem{Carullo:2023gtf}
G.~Carullo, R.~Cotesta, E.~Berti, and V.~Cardoso, ``{Reply to Comment on ''Analysis of Ringdown Overtones in GW150914''},'' \href{http://dx.doi.org/10.1103/PhysRevLett.131.169002}{{\em Phys. Rev. Lett.} {\bfseries 131} (2023) 169002}, \href{http://arxiv.org/abs/2310.20625}{{\ttfamily arXiv:2310.20625 [gr-qc]}}.

\bibitem{Isi:2023nif}
M.~Isi and W.~M. Farr, ``{Comment on \textquotedblleft{}Analysis of Ringdown Overtones in GW150914\textquotedblright{}},'' \href{http://dx.doi.org/10.1103/PhysRevLett.131.169001}{{\em Phys. Rev. Lett.} {\bfseries 131} no.~16, (2023) 169001}, \href{http://arxiv.org/abs/2310.13869}{{\ttfamily arXiv:2310.13869 [astro-ph.HE]}}.

\bibitem{Cotesta:2022pci}
R.~Cotesta, G.~Carullo, E.~Berti, and V.~Cardoso, ``{Analysis of Ringdown Overtones in GW150914},'' \href{http://dx.doi.org/10.1103/PhysRevLett.129.111102}{{\em Phys. Rev. Lett.} {\bfseries 129} no.~11, (2022) 111102}, \href{http://arxiv.org/abs/2201.00822}{{\ttfamily arXiv:2201.00822 [gr-qc]}}.

\bibitem{Baibhav:2023clw}
V.~Baibhav, M.~H.-Y. Cheung, E.~Berti, V.~Cardoso, G.~Carullo, R.~Cotesta, W.~Del~Pozzo, and F.~Duque, ``{Agnostic black hole spectroscopy: Quasinormal mode content of numerical relativity waveforms and limits of validity of linear perturbation theory},'' \href{http://dx.doi.org/10.1103/PhysRevD.108.104020}{{\em Phys. Rev. D} {\bfseries 108} no.~10, (2023) 104020}, \href{http://arxiv.org/abs/2302.03050}{{\ttfamily arXiv:2302.03050 [gr-qc]}}.

\bibitem{Bucciotti:2023ets}
B.~Bucciotti, A.~Kuntz, F.~Serra, and E.~Trincherini, ``{Nonlinear quasi-normal modes: uniform approximation},'' \href{http://dx.doi.org/10.1007/JHEP12(2023)048}{{\em JHEP} {\bfseries 12} (2023) 048}, \href{http://arxiv.org/abs/2309.08501}{{\ttfamily arXiv:2309.08501 [hep-th]}}.

\bibitem{Perrone:2023jzq}
D.~Perrone, T.~Barreira, A.~Kehagias, and A.~Riotto, ``{Non-linear black hole ringdowns: An analytical approach},'' \href{http://dx.doi.org/10.1016/j.nuclphysb.2023.116432}{{\em Nucl. Phys. B} {\bfseries 999} (2024) 116432}, \href{http://arxiv.org/abs/2308.15886}{{\ttfamily arXiv:2308.15886 [gr-qc]}}.

\bibitem{Bucciotti:2024zyp}
B.~Bucciotti, L.~Juliano, A.~Kuntz, and E.~Trincherini, ``{Quadratic Quasi-Normal Modes of a Schwarzschild Black Hole},'' \href{http://arxiv.org/abs/2405.06012}{{\ttfamily arXiv:2405.06012 [gr-qc]}}.

\bibitem{Isi:2021iql}
M.~Isi and W.~M. Farr, ``{Analyzing black-hole ringdowns},'' \href{http://arxiv.org/abs/2107.05609}{{\ttfamily arXiv:2107.05609 [gr-qc]}}.

\bibitem{csvQuadratic}
``\url{https://github.com/akuntz00/QuadraticQNM}.''.

\bibitem{Dhani:2020nik}
A.~Dhani, ``{Importance of mirror modes in binary black hole ringdown waveform},'' \href{http://dx.doi.org/10.1103/PhysRevD.103.104048}{{\em Phys. Rev. D} {\bfseries 103} no.~10, (2021) 104048}, \href{http://arxiv.org/abs/2010.08602}{{\ttfamily arXiv:2010.08602 [gr-qc]}}.

\bibitem{1961RSPSA.264..309S}
R.~{Sachs}, ``{Gravitational Waves in General Relativity. VI. The Outgoing Radiation Condition},'' \href{http://dx.doi.org/10.1098/rspa.1961.0202}{{\em Proceedings of the Royal Society of London Series A} {\bfseries 264} no.~1318, (Nov., 1961) 309--338}.

\bibitem{1962RSPSA.270..103S}
R.~K. {Sachs}, ``{Gravitational Waves in General Relativity. VIII. Waves in Asymptotically Flat Space-Time},'' \href{http://dx.doi.org/10.1098/rspa.1962.0206}{{\em Proceedings of the Royal Society of London Series A} {\bfseries 270} no.~1340, (Oct., 1962) 103--126}.

\bibitem{1962JMP.....3..566N}
E.~{Newman} and R.~{Penrose}, ``{An Approach to Gravitational Radiation by a Method of Spin Coefficients},'' \href{http://dx.doi.org/10.1063/1.1724257}{{\em Journal of Mathematical Physics} {\bfseries 3} no.~3, (May, 1962) 566--578}.

\bibitem{Hui:2020xxx}
L.~Hui, A.~Joyce, R.~Penco, L.~Santoni, and A.~R. Solomon, ``{Static response and Love numbers of Schwarzschild black holes},'' \href{http://dx.doi.org/10.1088/1475-7516/2021/04/052}{{\em JCAP} {\bfseries 04} (2021) 052}, \href{http://arxiv.org/abs/2010.00593}{{\ttfamily arXiv:2010.00593 [hep-th]}}.

\bibitem{1975RSPSA.344..441C}
S.~{Chandrasekhar} and S.~{Detweiler}, ``{The Quasi-Normal Modes of the Schwarzschild Black Hole},'' \href{http://dx.doi.org/10.1098/rspa.1975.0112}{{\em Proceedings of the Royal Society of London Series A} {\bfseries 344} no.~1639, (Aug., 1975) 441--452}.

\bibitem{PhysRevD.34.384}
E.~W. Leaver, ``Spectral decomposition of the perturbation response of the schwarzschild geometry,'' \href{http://dx.doi.org/10.1103/PhysRevD.34.384}{{\em Phys. Rev. D} {\bfseries 34} (Jul, 1986) 384--408}. \url{https://link.aps.org/doi/10.1103/PhysRevD.34.384}.

\bibitem{Silva:2023cer}
H.~O. Silva, G.~Tambalo, K.~Glampedakis, and K.~Yagi, ``{Gravitational radiation from a particle plunging into a Schwarzschild black hole: Frequency-domain and semirelativistic analyses},'' \href{http://dx.doi.org/10.1103/PhysRevD.109.024036}{{\em Phys. Rev. D} {\bfseries 109} no.~2, (2024) 024036}, \href{http://arxiv.org/abs/2308.14823}{{\ttfamily arXiv:2308.14823 [gr-qc]}}.

\bibitem{PhysRevD.35.3621}
S.~Iyer and C.~M. Will, ``Black-hole normal modes: A wkb approach. i. foundations and application of a higher-order wkb analysis of potential-barrier scattering,'' \href{http://dx.doi.org/10.1103/PhysRevD.35.3621}{{\em Phys. Rev. D} {\bfseries 35} (Jun, 1987) 3621--3631}. \url{https://link.aps.org/doi/10.1103/PhysRevD.35.3621}.

\bibitem{Iyer:1986vv}
S.~Iyer and C.~M. Will, ``{BLACK HOLE NORMAL MODES: A SEMIANALYTIC APPROACH. 1. FOUNDATIONS},''.

\bibitem{1985ApJ...291L..33S}
B.~F. {Schutz} and C.~M. {Will}, ``{Black hole normal modes - A semianalytic approach},'' \href{http://dx.doi.org/10.1086/184453}{{\em \apjl} {\bfseries 291} (Apr., 1985) L33--L36}.

\bibitem{2003PhRvD..68b4018K}
R.~A. {Konoplya}, ``{Quasinormal behavior of the D-dimensional Schwarzschild black hole and the higher order WKB approach},'' \href{http://dx.doi.org/10.1103/PhysRevD.68.024018}{{\em \prd} {\bfseries 68} no.~2, (July, 2003) 024018}, \href{http://arxiv.org/abs/gr-qc/0303052}{{\ttfamily arXiv:gr-qc/0303052 [gr-qc]}}.

\bibitem{Matyjasek_2017}
J.~Matyjasek and M.~Opala, ``Quasinormal modes of black holes: The improved semianalytic approach,'' \href{http://dx.doi.org/10.1103/physrevd.96.024011}{{\em Physical Review D} {\bfseries 96} no.~2, (Jul, 2017) }. \url{https://doi.org/10.1103%2Fphysrevd.96.024011}.

\bibitem{Hatsuda:2023geo}
Y.~Hatsuda and M.~Kimura, ``{Perturbative quasinormal mode frequencies},'' \href{http://arxiv.org/abs/2307.16626}{{\ttfamily arXiv:2307.16626 [gr-qc]}}.

\bibitem{motl2003asymptotic}
L.~Motl and A.~Neitzke, ``Asymptotic black hole quasinormal frequencies,'' 2003.

\bibitem{Ansorg:2016ztf}
M.~Ansorg and R.~Panosso~Macedo, ``{Spectral decomposition of black-hole perturbations on hyperboloidal slices},'' \href{http://dx.doi.org/10.1103/PhysRevD.93.124016}{{\em Phys. Rev. D} {\bfseries 93} no.~12, (2016) 124016}, \href{http://arxiv.org/abs/1604.02261}{{\ttfamily arXiv:1604.02261 [gr-qc]}}.

\bibitem{Ripley:2022ypi}
J.~L. Ripley, ``{Computing the quasinormal modes and eigenfunctions for the Teukolsky equation using horizon penetrating, hyperboloidally compactified coordinates},'' \href{http://dx.doi.org/10.1088/1361-6382/ac776d}{{\em Class. Quant. Grav.} {\bfseries 39} no.~14, (2022) 145009}, \href{http://arxiv.org/abs/2202.03837}{{\ttfamily arXiv:2202.03837 [gr-qc]}}.

\bibitem{Aminov:2020yma}
G.~Aminov, A.~Grassi, and Y.~Hatsuda, ``{Black Hole Quasinormal Modes and Seiberg\textendash{}Witten Theory},'' \href{http://dx.doi.org/10.1007/s00023-021-01137-x}{{\em Annales Henri Poincare} {\bfseries 23} no.~6, (2022) 1951--1977}, \href{http://arxiv.org/abs/2006.06111}{{\ttfamily arXiv:2006.06111 [hep-th]}}.

\bibitem{Bonelli:2021uvf}
G.~Bonelli, C.~Iossa, D.~P. Lichtig, and A.~Tanzini, ``{Exact solution of Kerr black hole perturbations via CFT2 and instanton counting: Greybody factor, quasinormal modes, and Love numbers},'' \href{http://dx.doi.org/10.1103/PhysRevD.105.044047}{{\em Phys. Rev. D} {\bfseries 105} no.~4, (2022) 044047}, \href{http://arxiv.org/abs/2105.04483}{{\ttfamily arXiv:2105.04483 [hep-th]}}.

\bibitem{Bonelli:2022ten}
G.~Bonelli, C.~Iossa, D.~Panea~Lichtig, and A.~Tanzini, ``{Irregular Liouville Correlators and Connection Formulae for Heun Functions},'' \href{http://dx.doi.org/10.1007/s00220-022-04497-5}{{\em Commun. Math. Phys.} {\bfseries 397} no.~2, (2023) 635--727}, \href{http://arxiv.org/abs/2201.04491}{{\ttfamily arXiv:2201.04491 [hep-th]}}.

\bibitem{Aminov:2023jve}
G.~Aminov, P.~Arnaudo, G.~Bonelli, A.~Grassi, and A.~Tanzini, ``{Black hole perturbation theory and multiple polylogarithms},'' \href{http://arxiv.org/abs/2307.10141}{{\ttfamily arXiv:2307.10141 [hep-th]}}.

\bibitem{leaver}
E.~W. Leaver, ``An analytic representation for the quasi-normal modes of kerr black holes,'' {\em Proceedings of the Royal Society of London. Series A, Mathematical and Physical Sciences} {\bfseries 402} no.~1823, (1985) 285--298. \url{http://www.jstor.org/stable/2397876}.

\bibitem{Lagos_2023}
M.~Lagos and L.~Hui, ``Generation and propagation of nonlinear quasinormal modes of a schwarzschild black hole,'' \href{http://dx.doi.org/10.1103/physrevd.107.044040}{{\em Physical Review D} {\bfseries 107} no.~4, (Feb, 2023) }. \url{https://doi.org/10.1103%2Fphysrevd.107.044040}.

\bibitem{Maggiore:1900zz}
M.~Maggiore, {\em {Gravitational Waves. Vol. 1: Theory and Experiments}}.
\newblock Oxford Master Series in Physics. Oxford University Press, 2007.
\newblock
\url{http://www.oup.com/uk/catalogue/?ci=9780198570745}.
\newblock

\bibitem{herdeiro2015asymptotically}
C.~A.~R. Herdeiro and E.~Radu, ``Asymptotically flat black holes with scalar hair: a review,'' 2015.

\bibitem{Bertone:2019irm}
G.~Bertone {\em et~al.}, ``{Gravitational wave probes of dark matter: challenges and opportunities},'' \href{http://dx.doi.org/10.21468/SciPostPhysCore.3.2.007}{{\em SciPost Phys. Core} {\bfseries 3} (2020) 007}, \href{http://arxiv.org/abs/1907.10610}{{\ttfamily arXiv:1907.10610 [astro-ph.CO]}}.

\bibitem{May:2024rrg}
T.~May, S.~Ma, J.~L. Ripley, and W.~E. East, ``{Nonlinear effect of absorption on the ringdown of a spinning black hole},'' \href{http://arxiv.org/abs/2405.18303}{{\ttfamily arXiv:2405.18303 [gr-qc]}}.

\bibitem{BHPToolkit}
``{Black Hole Perturbation Toolkit}.'' (\href{http://bhptoolkit.org/}{bhptoolkit.org}).

\end{thebibliography}\endgroup

\end{document}